\documentclass[lettersize,journal]{IEEEtran}
\usepackage{amsmath,amsfonts}
\usepackage{algorithmic}
\usepackage{algorithm}
\usepackage{array}
\usepackage[caption=false,font=footnotesize]{subfig}
\usepackage{textcomp}
\usepackage{stfloats}
\usepackage{url}
\usepackage{verbatim}
\usepackage{graphicx}
\usepackage{cite}
\usepackage{xcolor}
\usepackage{bm}
\usepackage{booktabs} 
\usepackage{multirow}

\usepackage{tikz}
\usetikzlibrary{shadows,shapes,arrows}
\usepackage{pgfplots}
\usetikzlibrary{pgfplots.groupplots}
\pgfplotsset{compat=1.18} 

\usepackage[normalem]{ulem}

\DeclareMathAlphabet{\mathcal}{OMS}{cmsy}{m}{n}
\SetMathAlphabet{\mathcal}{bold}{OMS}{cmsy}{b}{n}

\newcommand{\rulesep}{\unskip\ \vrule width 0.01pt\ }

\usepackage{amsmath,amssymb,amsfonts}
\usepackage{algorithmic}
\usepackage{graphicx}
\usepackage{textcomp}
\allowdisplaybreaks
\usepackage{bm}
\usepackage{mathtools}
\usepackage{xfrac}
\usepackage{multirow}

\usepackage[margin=0.57in]{geometry}

\DeclarePairedDelimiter\abs{\lVert}{\Vert}
\DeclarePairedDelimiter\norm{\lvert}{\vert}

\begin{document}

\title{Graph Neural Networks for Fast {Contingency} Analysis of Power Systems}

\author{Agnes M. Nakiganda,~\IEEEmembership{Member,~IEEE,} Spyros Chatzivasileiadis,~\IEEEmembership{Senior Member,~IEEE}
\thanks{A. Nakiganda is with the Department of Electrical and Electronic Engineering at Imperial College London, London, UK (a.nakiganda@imperial.ac.uk).}%
\thanks{S. Chatzivasileiadis is with the Department of Wind and Energy Systems at the Technical University of Denmark (DTU), Lyngby, Denmark.}%
\thanks{This work is supported by the SYNERGIES project, funded by the European Commission Horizon Europe program, Grant Agreement No. 101069839, and by the European Research Council (ERC) Starting Grant VeriPhIED, Grant Agreement No. 949899.}
}



\maketitle

\begin{abstract}
The successful integration of machine learning models into decision support tools for grid operation hinges on effectively capturing the topological changes in daily operations. Frequent grid reconfigurations and \mbox{N-$k$} security analyses have to be conducted to ensure a reliable and secure power grid, leading to a vast combinatorial space of possible topologies and operating states. This combinatorial complexity, which increases with grid size, poses a significant computational challenge for traditional solvers. In this paper, we combine Physics-Informed Neural Networks with graph-aware neural network architectures, i.e., a Guided-Dropout (GD) and an Edge-Varying Graph Neural Network (GNN) architecture to learn the set points for a grid that considers all probable single-line reconfigurations (all critical N$-1$ scenarios) and subsequently apply the trained models to \mbox{N-$k$} scenarios. We demonstrate how incorporating the underlying physical equations for the network equations within the training procedure of the GD and the GNN architectures performs with N$-1$, N$-2$, and N$-3$ case studies.  
Using the AC Power Flow as a guiding application, we test our methods on the 6-bus, 24-bus, 57-bus, and 118-bus systems. We find that GNN not only achieves the task of contingency screening with satisfactory accuracy but does this up to 400 times faster than the Newton-Raphson power flow solver. Moreover, our results provide a comparison of the GD and GNN models in terms of accuracy and computational speed and provide recommendations on their adoption for contingency analysis of power systems.
\end{abstract}


\section{Introduction}
\IEEEPARstart{T}{he} combinatorial problem arising from grid reconfigurations, topology optimization, and N$-k$ security topology changes presents a significant challenge in power grid management. It is vital that system operators can ascertain that potentially critical contingency scenarios are promptly screened and analyzed and that mitigation measures are devised. Power systems today are designed with inherent N$-1$ operational reliability; however, the likelihood of failure of multiple elements is high, especially due to the interconnected nature of the grid as more devices are being integrated and it becomes smarter \cite{4483749}. If not sufficiently addressed, such multiple contingencies can result in voltage collapse and cascading failures \cite{6152191,8691583}. 

Traditionally, numerical methods such as Newton-Raphson have served as a means to solve the power flow problem for critical contingency screening analysis. However, the computing time necessary for handling the combinatorial explosion of N$-k$  scenarios with such techniques becomes prohibitive as the system size increases. For instance, for the 118-bus\cite{PGLib} test case, while we only need to assess 166 N$-1$ potential topological changes, when it comes to N$-3$, we need to assess over 790'000 potential topologies, and this number considers only line disconnections and a single operating point. In other words, if solving an AC Power Flow for the 118-bus takes 1 second, we will complete an N$-1$ security analysis in just 3 minutes, but we will need more than \emph{9 full days} to complete a N$-3$ analysis with a single computer.
In Table \ref{tab:nw params}, we present an overview of the number of topologies that need to be captured for different networks for every single operating point.

\begin{table}[!b]
    \vspace{-2em}
    \centering
    \caption{Characteristics of the Test Networks}
    \resizebox{.90\linewidth}{!}
{
    \begin{tabular}{cccccccc}
        \toprule
        \text{Network} & \text{6-Bus}& \text{24-bus} & \text{57-bus} & \text{118-bus} \\
        \midrule
        Nodes & 6 & 24 & 57 & 118\\
        Branches &11  & 33 &  63 & 173\\
        Transformers & 0 & 5 & 17 & 13\\
        Generators &  2&10 & 6 & 53\\
        Loads & 3 & 17 & 42 & 99\\
        Eligible N$-1$ topologies & 11 & 32 & 62 & 166\\
        Eligible N$-2$ topologies & 55 & 505 & 1'928 & 14'408\\
        Eligible N$-3$ topologies &165  & 4'885 & 37'765 & 793'206\\
        \bottomrule
    \end{tabular}}\label{tab:nw params} \vspace{-0.8em} 
\end{table}

Machine Learning (ML) models, for example, Decision Trees, Support Vector Machines, Random Forests, and Neural Networks have been shown to handle complex power system problems tractably and efficiently \cite{9844246,9091534}. These models, once trained, eliminate the computationally intensive iterative procedures of traditional power flow solvers and scale well with increasing grid sizes. However, the downside to many ML algorithms is that they are often unable to consider grid topologies beyond the one topology on which they have been trained, i.e., they do not incorporate variables that relate to the connection/disconnection of power lines or the reconfiguration of buses. Moreover, training a single model for each potential N-$1$ topology would also be impractical. Their inability to capture different topologies is an inherent aspect of the contingency assessment problem in power systems, which hinders their adoption in real systems.

To leverage the enormous computational efficiency available with ML methods for application to the N-$k$ power flow problem, various ML-based architectures have been proposed. In \cite{guided2018}, a one-hot encoding that adds extra binary variables to represent the connection/disconnection of components was presented. However, results therein show this method may not scale well to larger systems with hundreds of components. In \cite{guided2018} and \cite{donnot-01906170, DONON2020316}, the authors introduce the so-called "Guided Dropout" method to address the topology change problem.
``Guided Dropout'' \textit{sparsifies} a neural network by introducing conditional neurons for each topology configuration, a feature that is suitable for addressing grid topology changes during {N$-k$} scenarios. Another NN-based architecture suitable for topology changes is the Graph Neural Networks (GNNs). Initially presented in \cite{4700287}, GNNs extend traditional NNs to data represented in graph domains by inherently exploiting the system structural characteristics, enabling the inclusion of information related to both graph nodes (buses) and graph edges (transmission lines) at both input and output of a model. GNNs have been applied to find a solution to the Optimal Power Flow problem in \cite{9053140,liu2022topology,falconer2022leveraging,9962773,8851855}. 

While these methods all show promising results, there is a lack of a detailed analysis to exploit their ability to truly generalize to topologies they have never seen before.
 For example, in the existing literature so far, they are trained for N$-1$ scenarios and then tested on a different set of N-1 scenarios and have to be re-trained to accommodate each possible N$-k$ scenario. Such a solution to retrain the GNN with new topologies has been proposed in \cite{9992121}. However, this is not very effective given the combinatorial number of topologies and associated scenarios for even medium-sized networks. The true advantage would be realized if they are trained for N$-1$ scenarios and can generalize to any N$-2$, or N$-3$, etc., scenarios without having seen any of those topologies before. In the case of the 118-bus system, this means training the ML model with only 166 N$-1$ topologies and having the ability to screen over 790'000 N$-3$ topologies for various operational scenarios. The savings in computational time in the latter case will be substantial.
 
This paper therefore investigates the generalization capabilities of two topology-aware NN architectures i.e., Graph Neural Networks and Guided Neural Networks, trained on solely the reference case (N$-0$) and N$-1$ topologies, and tested on unseen N$-2$ and N$-3$ topologies, to enable faster contingency screening based on the AC power flow model. This eliminates the requirement to generate training data for N$-2$ topologies (with an $O(N^2)$ complexity) and N$-3$ topologies (with a $O(N^3)$ complexity), resulting in huge computational gains. The contributions of this paper are three-fold:
\begin{itemize}
    \item First, we introduce physical equations into the training phase for each of the NN architectures in the form of the nodal power balance equations, hereafter referred to as Physics-Informed Graph Neural Networks (PI-GNN) and Physics-Informed Guided Dropout Neural Networks (PI-GDNN). This is done to enhance the feasibility of the algorithms in a physical network. Moreover, by including collocation data points trained only on the physical equations, we can reduce the training data requirements of the solution framework.
    \item Second, we compare the accuracy and scalability against the traditional Newton-Raphson power flow solver and DC power flow approximation using data sets derived from test networks of grid sizes 6-bus, 24-bus, 57-bus, and 118-bus for various \text{N$-1$} scenarios. 
    \item Third, we analyze how well each of the models generalizes to scenarios of network topologies generated from \text{N$-2$} and \text{N$-3$} without the requirement for re-training the model thus providing conclusions on their applicability to \text{N$-k$} instantiations.
\end{itemize}

The rest of this paper is structured as follows: Section \ref{sec:model} describes the topology-aware NN models GD and GNN adopted in this work while Section \ref{sec:framework} presents the modelling considerations and solution methodology proposed. We demonstrate and discuss the results from case studies in Section~\ref{sec:results}, and provide concluding remarks in Section~\ref{sec:concl}.

\section{Graph-Aware Neural Network Models} \label{sec:model}
A traditional feed-forward NN includes linear transformations and non-linear activations at various layers of the network i.e.
\begin{subequations} \label{eqn: NN layer}
    \begin{align}
       &{\bm{z}}_{0} = \bm{x},\\
        &{\bm{z}}_{l+1} = \sigma({\bm{W}_{l+1}}\bm{z}_{l}+{\bm{b}_{l+1}}) &\forall l = 1,\dots,H,\label{NN1}\\
        & \hat{\bm{y}} = {\bm{W}_{H+1}}\bm{z}_{H}+{\bm{b}_{H+1}},
    \end{align}
\end{subequations}
where $\bm{x}$ is the input, ${\bm{z}}_{l}$ is the output of the neurons in layer $l$, $\bm{W}_{l+1} \in \mathbb{R}^{\mathrm{F_i}\times \mathrm{F_o}}$ and $\bm{b}_{l+1} \in \mathbb{R}^{F\mathrm{o}}$ denote the learnable weights and biases connecting layer $l$ and $l+1$. Parameters ${\mathrm{F_i}}$ and ${\mathrm{F_o}}$ denote the number of input and output features, respectively. The function $\sigma(\cdot)$ is the non-linear activation function applied at each layer of the NN, $\hat{\bm{y}} $ denotes the predictions of the NN model, while $H$ is the total number of layers included in the NN model.  

During the training process that maps $\bm z_0 \mapsto \hat{\bm y}$, the goal is to minimize a loss function $\mathcal{L}$ by adjusting the weights and biases. Given a training dataset $\mathcal{D}_t$, we define the loss using the Mean Square Error (MSE) between the prediction $\hat{\bm y}$ and the ground truth $\bm y$ as: 
\begin{align}\label{eqn: loss}
    \mathcal{L} =  \frac{1}{\mathcal{D}_t} \sum_{i=1}^{\norm{\mathcal{D}_t}}  {\abs{\bm{\hat{y}} -\bm{y}}}^2
\end{align}

\vspace{-0.5em}
\subsection{Guided Dropout Neural Network (GDNN)}
  The Guided Dropout NN technique proposed in \cite{donnot-01906170} asserts that it is possible to train only one network with variants that capture all the different N$-1$ topologies around a standard reference topology; in the reference topology, all lines are in service. The reference topology and all other N-$1$ cases are trained simultaneously, wherein the N$-1$ cases have been encoded in the NN by activating ``conditional'' $N^{\mathrm{GD}}$ hidden units (i.e. neurons) in layer $l \in \mathcal{H}$. Specifically, a number of static neurons $N^{\mathrm{S}}_{l}$ are applied in the $H$ layers, and these neurons are set to always be active.  Another set $N^{\mathrm{GD}}_l$ is activated based on the considered N-$1$ topology; for example, a different single guided neuron can represent each of the $N^{\mathrm{N-1}}$ configurations defining the N$-1$ scenarios. These are subsequently added to $l \in \mathcal{H}$ to represent all configurations that move away from the reference. In this work, the reference refers to the nominal power grid topology that does not have any buses or lines disconnected, i.e., N$-0$ case.  The total number of neurons per layer is thus defined as: $N_l = N^{\mathrm{S}}_{l} + N^{\mathrm{GD}}_l$ where $N^{\mathrm{GD}}_l \geq \norm{\mathcal{E}}$ (where $\norm{\mathcal{E}}$ is the number of edges, $e$, in the power network). Note that it is possible to have $N^{\mathrm{GD}} = 0$ in some layers; this is normally the case for the first and last layers in the model.

  A binary structural vector $\bm\tau \in \{0,1 \}^{N_l}$ encoded with each N$-1$ case, is used to guide the ``drop-in'' of neurons defined as:
  \begin{align}\label{eqn:gd_str}
      &\hspace{3cm} \bm\tau \in \mathbb{B}^{N_l} \\
      &\text{where} \nonumber \\
      &{N_l}  = N^{\mathrm{S}}_{l} + N^{\mathrm{GD}}_l = N^{\mathrm{S}}_{l} +   N^{\mathrm{N-1}} = N^{\mathrm{S}}_{l} +\alpha\cdot\norm{\mathcal{E}} \\
    &\bm\tau = \{\bm\tau_s , \bm\tau_e \} = 
    \begin{cases}
     \tau_s = 1 & \text{if } s=\{1, \cdots, N^{\mathrm{S}}_{l}\} \\
  \tau_e = 0 & \text{if } e =\{ 1, \cdots, N^{\mathrm{N-1}}\} \,\land\, e = \text{``on''} \\
 \tau_e = 1 & \text{if } e =\{ 1, \cdots, N^{\mathrm{N-1}}\} \,\land\, e = \text{``off''}
    \end{cases}
  \end{align}  
   where $\alpha \geq 1$ is a factor that varies the number of conditional neurons ``dropped in'' per edge $e$, $\bm\tau_s$ and $\bm\tau_e $ are vectors applied to the stationary and conditional neurons to incorporate the topological information in the NN. Section III discusses the choice of $\alpha$ in further detail. Vector $\bm \tau$ is passed through to the NN to initialize the masks and subsequently turns on the "conditional" neurons in one or more of the NN layers. The updates of the NN at each layer in \eqref{NN1} are thus modified as follows:
  \begin{align}
      &{\bm{z}}_{l+1} = \sigma({\bm{W}_{l+1}}\bm{z}_{l}\odot \tau +{\bm{b}_{l+1}}) &\forall l = 1,\dots,H\label{eqn: GD layer}
  \end{align}
  
\begin{figure}[bt!]
    \centering
    \includegraphics[width = 0.95\linewidth]{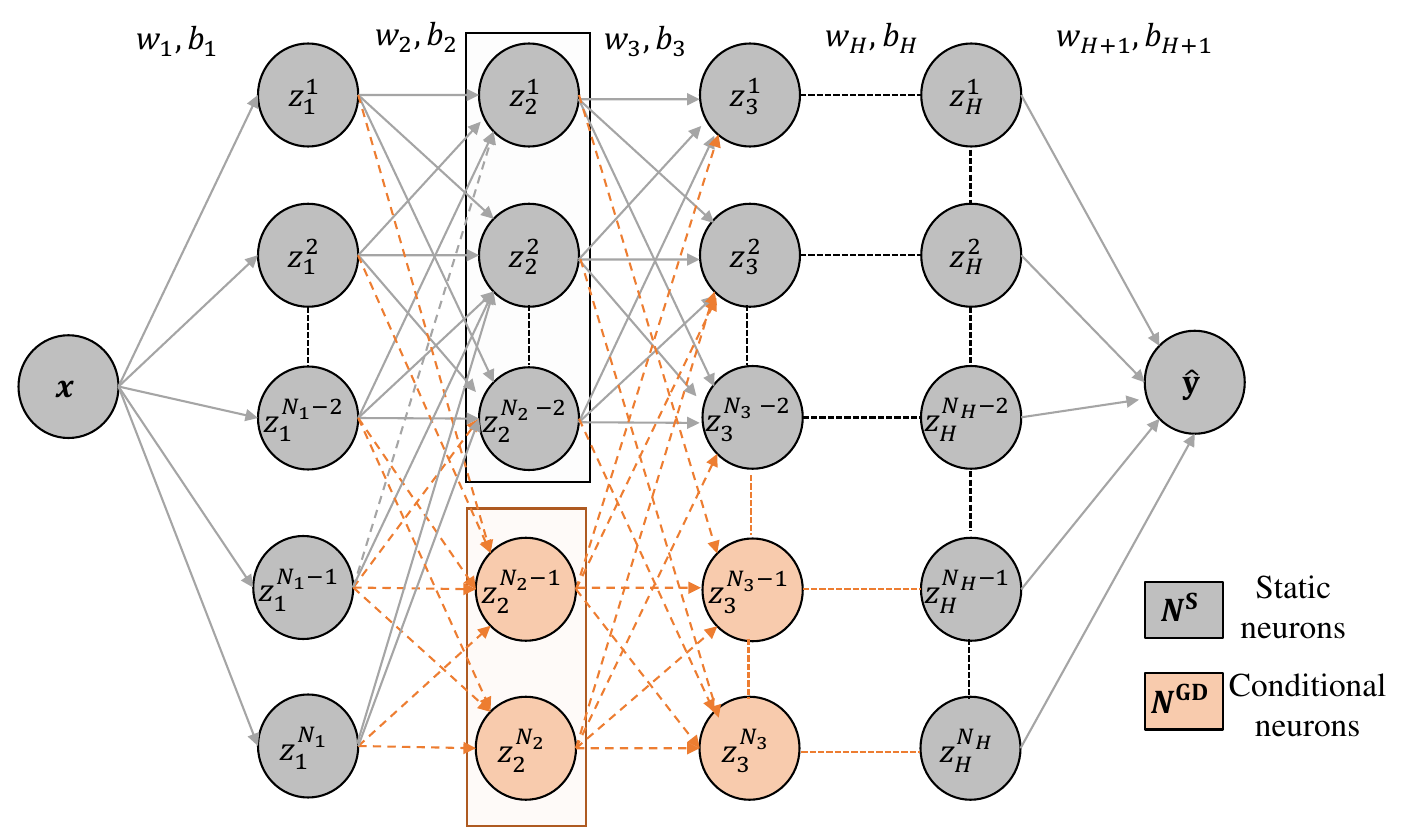}
    \caption{Illustration of the Guided Dropout technique on a NN architecture. One or more conditional neurons can be used to represent the different topologies of the grid.}
    \label{fig:standard}
    \vspace{-1.0em}
\end{figure}

  {Including more neurons for each N$-1$ topology can be done in two ways: either reducing the number of neurons $N^{\mathrm{S}}_l$, that capture the reference topology, while setting $\alpha = 1$ or including more $N^{\mathrm{GD}}_l$ i.e. $\alpha >1$.} 
Fig.~\ref{fig:standard} illustrates the architecture of the Guided Dropout NN.

\vspace{-1.0em}
\subsection{Graph Neural Networks (GNN)}
\begin{figure}[b!]

    \centering
    \includegraphics[width = 0.75\linewidth]{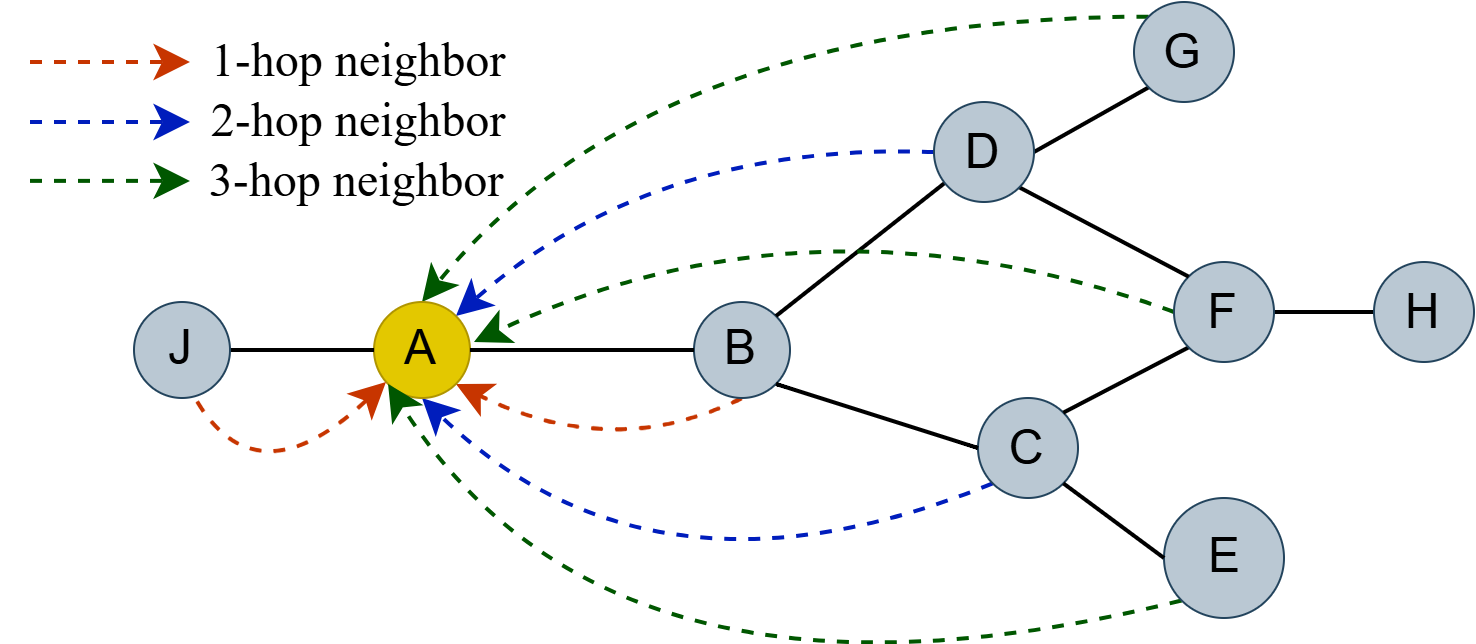}
    \caption{Illustration of the 1-hop, 2-hop, and 3-hop neighborhood shift of node A in the 9-bus network.}
    \label{fig:GNN_illustration}
    \vspace{-1.0em}
\end{figure}

The key property of the GNN is that it utilizes the message-passing framework \cite{gilmer2017neural}, where each node collects and aggregates information from its neighbors. Consider a graph $G = (\mathcal{V}, \mathcal{E})$, where node $i \in \mathcal{V}$ with $\norm{\mathcal{V}}=V$ and edge $e_{ij}=(i , j) \in \mathcal{E}$ with $\norm{\mathcal{E}}=E$. The neighborhood of a node is defined by $\mathcal{N}(i) = \{ j \in \mathcal{V} | (i,j) \in \mathcal{E}$. The information propagation at each node in every hidden layer follows:
\begin{itemize}
 \item Neighborhood feature computation:
        \begin{align}
            h_j^{k -1} | j\in \mathcal{N}(i) \label{eqn:gnn1}
        \end{align}
 \item Neighborhood feature aggregation at each node:
         \begin{align}
             a_i^{k} = \theta(h_j^{k -1} | j\in \mathcal{N}(i)) \label{eqn:gnn2}
         \end{align}
where $\theta$ is the aggregation function chosen.\\
 \item Node status update: 
             \begin{align}
                 h_i^{k} =\sigma(h_i^{k -1}, a_i^{k} ) \label{eqn:gnn3}
             \end{align}       
\end{itemize}
A Graph Convolution Neural Network (GCNN) is a GNN framework widely used to perform the message propagation task defined above by exploiting the convolution of graph signals \cite{7352352,8579589}. Consider a Graph Shift Operator (GSO), $\bm S \in \mathbb{R}^{V\times V}$, a matrix where $S_{ij}$ has non-zero entries if and only if $(i,j) \in \mathcal{E}$ or $i=j$, i.e. this can be the adjacency matrix of the network. Graph convolution involves the successive application of the GSO, $\bm S$, to the input graph signal, $x^{g}$, to create a sequence of graph shifted signals i.e., $X^{g}:=[ x^{g}, \bm{S}x^{g}, \ldots, \bm{S}^{K-1}x^{g}]$ where $K$ is the number of shifts. The result is the linear combination of features gathered from each node's neighborhood defined by the number of shifts. Figure \ref{fig:GNN_illustration} illustrates the neighborhoods of a given node in the network. 

A GCNN follows the message propagation framework by aggregating nodal information from its k-hop neighborhood, i.e., using successive communications with the one-hop neighbors. It is based on graph convolutions enabled with a shift operator matrix, $\bm S^{k}$, defined by the network's adjacency matrix, $A$. The propagation rule at each GCNN layer is defined as:\vspace{-0.5em}
\begin{align}
    & \bm Z^{}_{l} = \sigma \left(\sum_{k=0}^{K} {{\bm W^{}_{lk}}\bm S^{k}} {\bm Z^{}_{l -1}} + \bm b_{l} \right)\label{eqn: EV layer}
\end{align}\vspace{-0.2em}
The GCNN filter, $\sum_{k=0}^{K} {{\bm W^{}_{lk}}\bm S^{k}}$, applies the same weight (parameterization) $W^{}_{lk}$ per shift operation for all nodes i.e, weight $W^{}_{l1}$ will be used for all 1-hop, $W^{}_{l12}$, for all 2-hop e.t.c neighborhood information exchanges irrespective of the node and edges. The GCNN filter, therefore, weighs each shift operation (communication with the neighborhood) using the matrix of scalar parameters $W^{}_{lk}:=\{w_{l0}, \dots, w_{lK} \}$ whilst taking into account the underlying graph support using $S^{k}$. Note that since the scalar weights $\bm W^{}_{lk}$ are shared by all nodes for each neighborhood $k$, this characteristic means fewer parameters can be used to train the GCNN model. Parameter sharing while providing computational benefits limits the representational capacity of the GCNN due to its non-discriminatory nature, i.e. there is no distinction in the weights applied to information received from the different edges connected to a node.

A more generalized parameterization can be obtained by letting each node weigh each of its neighbors' information differently. This idea is presented in \cite{9536420} that defines an edge varying graph filter $\Phi \in \mathbb{R}^{V\times V}$ with the same sparsity pattern as $A$ i.e., $\Phi^{(k)}_{ij} \geq 0$ if $(i,j) \in \mathcal{E}$ and zero elsewhere. Unlike the Graph Convolution Network (GCNN) \cite{KipfW16} where neighborhood shift weights are shared among all nodes and edges, the edge varying graph filter $\Phi$ generalizes the GCNN by allowing the use of different trainable weights, $\Phi^{(k)}_{ij}$ per edge and node for each shift. This operation is described as follows: At each node $i$, the neighborhood features are computed as:
\begin{align}\label{eqn:GF1}
    z_{i}^{(k)} = \sum_{j\in \mathcal{N}(i)} \Phi^{(k)}_{ij} z_{j}^{(k)} 
\end{align}
 A sequence of features is gathered from multiple hops, $k$ (neighborhood shifts) as:
 \begin{align}\label{eqn:GF}
    \bm z_{}^{(k)} = {\bm \Phi^{(k:0)}} \bm x = \displaystyle \prod_{k'=0}^{k} \bm\Phi^{(k')} \bm x , \quad\forall k= 0, \cdots, K
\end{align}

The Edge-Varying GNN (EVGNN), which employs a graph convolution at its core to achieve the steps in the propagation model, is therefore defined as:

\begin{align}
    & \bm Z^{}_{l} = \sigma \left(\sum_{k=0}^{K} {\bm \Phi^{(k:0)}} {\bm S^{k} Z^{}_{l -1}} + \bm b_{l} \right)\label{eqn: EV layer2}
\end{align}
where the filter matrix is defined by the term $\sum_{k=0}^{K} {\bm \Phi^{(k:0)}}\bm S^{k}$ and edge weight matrix $\bm \Phi^{(k:0)}$ shares the same support of $\bm A + \bm I_{V}$. Matrix $\bm I$ is an identity diagonal matrix that ensures that the node's own input is included in the aggregation step at each layer. 
Note that while the EVGNN architecture eliminates the non-discriminatory nature of the GCNN, it requires a larger number of trainable parameters, and hence, its order of complexity increases with the number of nodes in the network.

In this work, we employ the more generalized form of the GCNN, the Edge-Varying GNN (EVGNN) model described in \eqref{eqn:GF1}-\eqref{eqn: EV layer2}. EVGNNs provide greater degrees of freedom by enabling node-specific and edge-specific parameterization for feature weighting from neighboring nodes. 
This characteristic ensures that more detail is captured during local neighborhood feature aggregation. tly reduce its generalization properties.

In the following, we describe the solution architecture used for the implementation of the aforementioned models for the prediction of nodal features, i.e. voltage and power generation during power system analysis.

\section{Solution Framework} \label{sec:framework}
We apply the two graph-aware NN models described in Section \ref{sec:model} to approximate the solution of the AC power flow problem. Our goal is to facilitate faster contingency screening and assessment. We consider three inputs, $\bm X \in \mathbb{R}^{3\times V}$ for the EVGNN and $\bm x \in \mathbb{R}^{3V}$ for the GDNN, these include the active and reactive power consumption, $P_l$ and $Q_l$, and the active power generation $P_g$ i.e. $ \bm x_i=: [\bm P_l, \bm Q_l, \bm P_g]$ at each node. We predict three outputs at each node which include the voltage magnitude $V$ and voltage angles $\delta$ at all PQ buses and the reactive power generation at the PV buses, i.e., $\hat{\bm y_i}=: [\bm V, \bm \delta, \bm Q_g]$ where $\hat{\bm Y} \in \mathbb{R}^{3\times V}$ is defined for the EVGNN and $\hat{\bm y} \in \mathbb{R}^{3V}$ for the GDNN.
In the following, we describe the pertinent architectural features for the GD and EVGNN models proposed in this work. 

\vspace{-1.0em}
\subsection{Modeling Considerations}
\subsubsection{Guided Dropout Neural Network Architecture}
\begin{itemize}
    \item As proposed in \cite{donnot-01906170}, the architecture includes three main building blocks, i.e., an encoder, GDNN layers, and a decoder. In this work, we use Fully-Connected NN (FCNN) layers for both the encoder and decoder, where the encoder transforms the latent variables into a low-dimensional subspace, and the decoder provides a function to compute the output. GDNN layers, \eqref{eqn: GD layer}, are applied within the encoder and decoder to propagate the features whilst applying the structural vector $\tau$, which is defined by the list of edges in the network.
    \item While not discussed in \cite{donnot-01906170}, the number of conditional neurons $N^{\mathrm{GD}}$ considered per line affects the accuracy of the GDNN architecture. Moreover, an increase in the number of neurons per line increases the number of parameters to be trained by the GDNN model.  Based on multiple sensitivity simulations, we determined that a minimum of two guided neurons per line is sufficient to effectively represent each topology change i.e., the length of vector $\tau_{\mathrm{e}}$ in \eqref{eqn:gd_str} is defined as $ N^{\mathrm{N-1}} = \alpha\cdot\norm{\mathcal{E}}=2\cdot\norm{\mathcal{E}}$. 
\end{itemize}

\subsubsection{Edge-varying GNN Architecture}
The EVGNN architecture similarly includes three building blocks.
\begin{itemize}
    \item  At the input, we include an encoder for each node to transform the nodal features into a low-dimension space, this is constructed using FCNN layers.
    \item The hidden layers consist of EVGNN layers described in \eqref{eqn: EV layer}. 
    \item A readout layer is included at the output of the EVGNN layers to enhance the globalization of the local aggregation performed at each node in the EVGNN layers, this layer is defined by a FCNN. 
\end{itemize}

\vspace{-1.0em}
\subsection{Incorporating the Network Physics}
A Physics-Informed Neural Network (PINN) extends the conventional NN by incorporating domain knowledge through the physical equations governing the system's behavior into the training procedure~\cite{9282004,RAISSI2019686}. This is achieved by modifying the loss function of the NN \eqref{eqn: loss} to include a regularization term defined as the ``physics-loss'', denoted here as $\epsilon$. Thus, the PINN can reduce over-fitting and learn a more generalizable model as compared to conventional NNs~\cite{RAISSI2019686}.

Considering a power network with nodal voltage magnitude $V_k$, voltage angle $\theta_{k}$, active and reactive generation, $P_{g,k}$ and $Q_{g,k}$, active and reactive demand, $P_{l,k}$ and $Q_{l,k}$, then the active and reactive power injections, $P_k$ and $Q_k$, can be obtained by: 
\begin{align} \label{eq:powerinj}
    & S_{k}  = P_{k} + \mathrm{j}Q_{k} = V_{k}e^{\mathrm{j}\delta_{k}}\left ( \bm Y_{k}V_{m}e^{\mathrm{j}\delta_{m}}\right)^{*}, 
\end{align} 
where $\bm Y = \bm G + \mathrm{j} \bm B$ is the nodal admittance matrix.

The power injections should satisfy the nodal power balance equation defined as:
\begin{align}\label{eq:nodal balance}
    & S_{g,k} -  S_{d,k}  - S_{k} = 0 
\end{align}

The expressions for active and reactive power balance at each node in \eqref{eq:nodal balance} can be represented in a compact form as:
\begin{align}
    f_P(\mathbf{G}, \mathbf{B}, \bm V_{}, \bm \delta_{}, {\bm P_{l}}, {\bm P_{g}}) = 0 \\
    f_Q(\mathbf{G}, \mathbf{B}, \bm V_{}, \bm \delta_{}, {\bm Q_{l}}, {\bm Q_{g}}) = 0
\end{align}

    To explicitly consider this behavior in the training of both the GDNN and the EVGNN models, we add a physics-loss of the form:
    \begin{align}
& \bm \epsilon_P =\max\left(0, \abs{f_P} - \gamma \right) \label{eq: p-err}  \\
&\bm\epsilon_Q =\max\left(0, \abs{f_Q} - \gamma\right) \label{eq: q-err}\\
&\bm\epsilon_{Q_{\mathrm{lim}}} = \max\left(0, (Q^{\mathrm{max}}_g - Q_g ), ( Q_g - Q^{\mathrm{min}}_g) \right) \label{eq: lim-err}
\end{align}

where \eqref{eq: p-err}, \eqref{eq: q-err}, and \eqref{eq: lim-err} relate to the errors in the active power balance, reactive power balance, and reactive generator limits at each node. The tolerance $\gamma$ corresponds to the precision with which we want the power balance equations to be respected.

As a result, the loss function presented in \eqref{eqn: loss} for training the NNs is amended to include the physics loss as:
\begin{subequations}
\begin{align}\label{eq:pinnloss}
&\hspace{5em}\mathcal{L} = \mathcal{L}_{0} + \mathcal{L}_{\bm\epsilon},\\
&\text{where}\nonumber\\
\begin{split}
   &\mathcal{L}_{0} =  \frac{1}{\mathcal{D}_t} \sum_{j=1}^{\norm{\mathcal{D}_t}} \left( \Lambda_v \abs{\bm{\hat{V}}^{j} -\bm{V}^{j}}^2 + \Lambda_\delta \abs{\bm{\hat{\delta}}^{j} -\bm{\delta}^{j}}^2\right .  \\ & \hspace{3cm} \left . + \Lambda_Q \abs{\bm{\hat{Q}_g}^{j} - \bm{Q_g}^{j}} ^2 \right)
\end{split}\\
&\mathcal{L}_{\bm\epsilon} = \frac{\Lambda_{\bm\epsilon}}{ \mathcal{D}_c} \sum_{j=1}^{\norm{\mathcal{D}_c}} \left({\abs{\bm\epsilon_P^{j}}}^2  +{\abs{\bm\epsilon_Q^{j}}}^2+ {\abs{\bm\epsilon_{Q_{\mathrm{lim}}}^{j}}}^2 \right) 
\end{align}    
\end{subequations}
$\norm{\mathcal{D}_t}$ denotes the number of training data points usually generated through simulation and $\norm{\mathcal{D}_c}$ is the number of collocation training points. Collocation points are randomly generated samples from the input space, $\Lambda_p, \Lambda_l$ and $\Lambda_{\epsilon}$ are hyper-parameters associated with each loss function. Their choice has a major influence on the performance of the PINN. 

\begin{figure}[!tb]
   \vspace{-1.0em}
       \centering
    \subfloat[PI-GDNN architecture \label{fig:PI-GDNN}]{%
     \includegraphics[width=.47\textwidth]{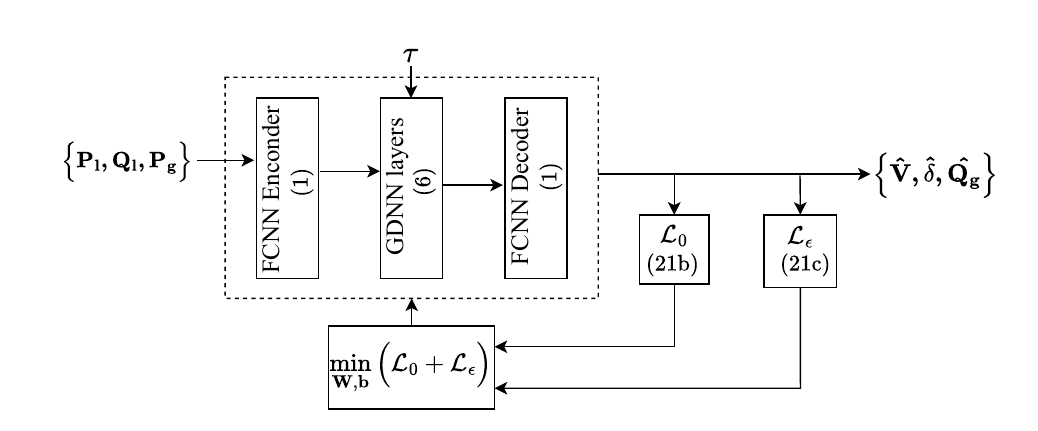}
    } 
    \hfill
    \vspace{-0.1em}
    \subfloat[PI-EVGNN architecture \label{fig:PI-EVGNN}]{%
    \includegraphics[width=.45\textwidth]{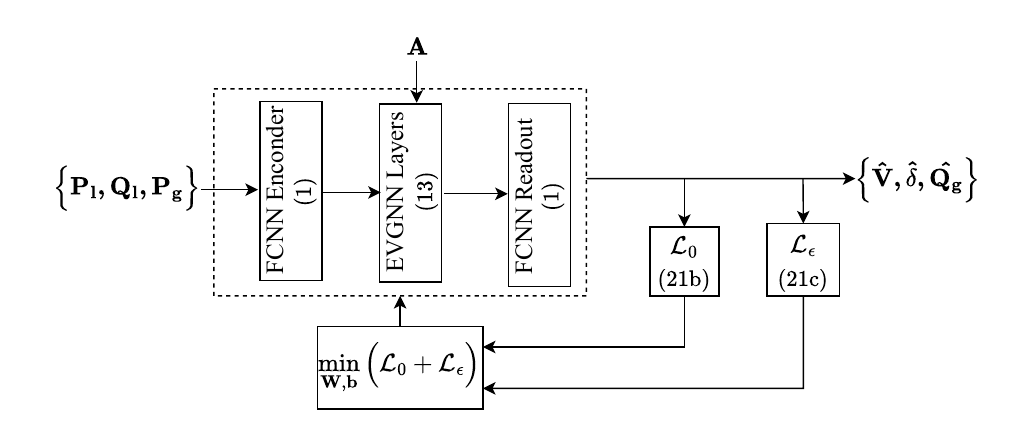}
    }

    \caption{Solution framework for the topology-aware PI-GDNN and PI-EVGNN architectures.} \vspace{-1em}
    \label{fig:arc} 
   \end{figure}

The physics loss described in \eqref{eq:pinnloss} is added to both the GDNN and EVGNN architectures therefore resulting in the Physics-Informed Guided Dropout Neural Network (PI-GDNN) Physics-Informed Edge Varying Graph Neural Network (PI-EVGNN). Figure \ref{fig:arc} presents an illustration of the solution architectures adopted for the PI-GDNN and PI-EVGNN frameworks. 

\section{Numerical Results and Discussion} \label{sec:results}
\subsection{Simulation Setup}
The GDNN, PI-GDNN, EVGNN and PI-EVGNN models presented in Section~\ref{sec:framework} are trained and tested to evaluate their accuracy in providing solutions to the AC power flow problem considering various operating states. Their performance is evaluated on the 6-bus, 24-bus, 57-bus, and 118-bus test systems from the PGLib-OPF network library v19.05\cite{PGLib}. The characteristics of the test networks are presented in Table \ref{tab:nw params}.  
{\color{black}
To evaluate the models, a dataset consisting of 1'000 load and generation scenarios is generated for each test network. To obtain the operating points, active and reactive power consumption and active power generation values at the nodes are uniformly sampled within a range of 60\% to 120\% of their nominal values in the reference grid. Demand and generation are varied independently for each scenario, and the slack bus covers potential generation-demand unbalance within its power limits. If a scenario is infeasible (i.e. we do not obtain a power flow solution), it is discarded, and a new scenario is sampled.
Each of these scenarios is simulated in the N$-0$ and for all eligible N$-1$ contingencies to obtain operating points. This means that for the 6-bus case, we generate 12'000 points [(1(N$-0$)+11(N$-1$) $\times$ 1000 scenarios], while for the 118-bus we generate 167'000 points. For the 24-bus and 57-bus, the reader can easily infer from Table~\ref{tab:nw params} that we generate 33'000 and 63'000 points, respectively. We use these points for the training, validation, and testing of the NNs. Of the operating points generated for each N$-0$ and N$-1$ case, 60\% is used as training data, and the rest are used for the validation and testing of the NNs.

The 1'000 generation and demand scenarios are then further simulated for the eligible N$-2$, and N$-3$ contingency cases. The operating points we obtain in this case are only used for the NN testing, i.e. to assess how well the NNs can estimate N$-2$ and N$-3$ violations; these data are \emph{not} used for training. Considering the thousands of possible N$-2$ and N$-3$ contingencies, we need to limit the number of topologies we consider to contain the necessary computing time to generate test scenarios. For the purposes of this study, we set a maximum of 500 eligible topologies for N$-2$ and 1'000 eligible topologies for N$-3$. This means that for each test system, we evaluate up to 1'500'000 data points (1'000 scenarios $\times$ (1'000 N$-2$ + 500 N$-3$) topologies), which will be used only for testing the generalization capabilities of the NNs. From these 1'500'000 data points for each test system, we keep only the ones for which the AC Power Flow converged. For example, contingencies that result in islanding, leaving parts of the network isolated, are discarded. {\color{black} This results in the following dataset size used for testing the NNs for each of the test systems: 89'405 and 637'795 for the 6-bus and 24-bus, respectively, and 616'632 and 346'478 for the 57-bus and 118-bus, respectively.
 }
We use Pandapower's \cite{8344496} power flow solver to simulate the operational scenarios for each of the operating points. The code and datasets are available online at \cite{git_GNN}.
}

The predictions of the nodal voltages, angles, and reactive power generation at each bus are provided by each of the models applied. We directly use these predictions to determine the real ($P_{km}$) and reactive ($Q_{km}$) power flows in the different lines based on the functions described in \eqref{eqn: lines}. The computational result in \eqref{eqn: lines} is then compared with the values obtained by simulations using the analytical power flow solver rather than predicting the power flows using an NN. 
\begin{subequations}\label{eqn: lines}
    \begin{align}
     & S_{km} =  V_{k}(I_{km})^*, \quad S_{mk} =  V_{m}(I_{mk})^*, \hspace{0.1cm} &\forall e_{km}\in \mathcal{E} \label{eqn:2a}\\
    & I_{km} = y_{km}^{s}(V_{k}-V_{m}) + y_{km}^{sh}V_{i} , & \forall e_{km}\in \mathcal{E} \label{eqn:3a}\\
    & I_{mk} = y_{km}^{s}(V_{m}-V_{k}) + y_{mk}^{sh}V_{m} , & \forall e_{km}\in \mathcal{E} \label{eqn:4a}
\end{align}
\end{subequations}
where $y^{s}_{km}$ is the series admittance while $y^{sh}_{km}$ is the shunt admittance of the line, respectively. The active and reactive power flows into the line at the sending (receiving) end are denoted by $S_{km} = P_{km} + {\mathbf{j}}Q_{km}$ ($S_{mk} = P_{mk} + {\mathbf{j}}Q_{mk}$). $I_{km}=|I_{km}|\angle \vartheta_{ij}$ ($I_{mk}=|I_{mk}|\angle \vartheta_{mk}$) is the current flowing into the line from sending (receiving) nodes.


The GDNN and PI-GDNN models are trained with three hidden layers and number of neurons per layer equal to \textcolor{black}{$N_{l}^{\mathrm{S}} = 10\times |\mathcal{V}|$ and $N_{l}^{\mathrm{GD}} =  2\times |\mathcal{E}|$}. The EVGNN and PI-EVGNN architectures include two hidden layers, and the neighborhood aggregation parameter is set as $K=2$. The PI-GDNN and PI-EVGNN models are trained with 50\% of the training scenarios used in the GDNN and EVGNN, while the other 50\% is replaced with collocation points. Table \ref{tab:trainparameters} presents the number of parameters used to train each of the models. The models are trained using Pytorch \cite{pytorch} with ADAM \cite{kingma2017adam} as the optimizer, the $\tanh$ activation function, and a learning rate set to 0.0005. The models were trained in a High-Performance Computing (HPC) server with an Intel Xeon E5-2650v4 processor and 256 GB RAM.

\begin{table}[!t]
\caption{Number of parameters used in training the GDNN and EVGNN models for the test networks.}\label{tab:trainparameters}
\renewcommand{\arraystretch}{1.2}
\resizebox{.99\linewidth}{!}
{
\begin{tabular}{cllrrrr}
\toprule
\multicolumn{2}{l}{Test Network}   &  & \multicolumn{1}{l}{6} & \multicolumn{1}{l}{24} & \multicolumn{1}{l}{57} & \multicolumn{1}{l}{118} \\ \midrule
 \multirow{2}{*}{\begin{tabular}[c]{@{}c@{}}Training\\ Parameters\end{tabular} } & \multirow{1}{*}{GD}  && 50'400 & 800'352 & 4'507'902 & 19'308'576 \\ \cline{2-7} 
 & \multirow{1}{*}{EVGNN}   &  & 34'992 & 208'864 & 1'011'358 & 4'065'872 \\
 \bottomrule
\end{tabular}}
\vspace{-1.5em}
\end{table}


\begin{figure}[!b]
\centering
\vspace{-1.0em}
     \scalebox{0.5}{ \includegraphics[width = 0.90\textwidth]{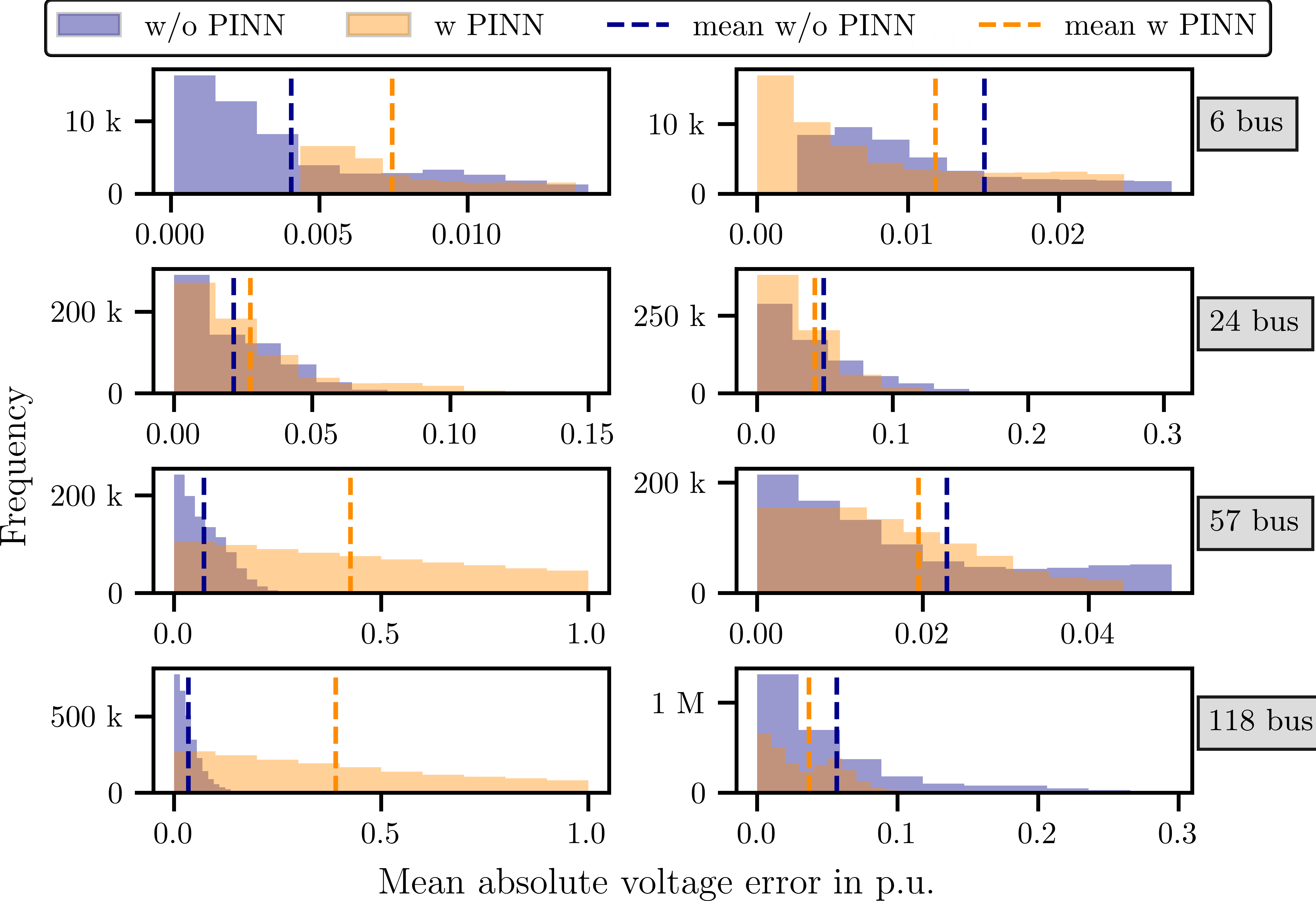}}
     \caption{Density distribution of the mean absolute error in voltage predictions for the GD-based and GNN-based models with and without physics included in the training under N-1 test scenarios. 
     {\footnotesize \it Note that different axes have been used for each plot to provide a clearer representation of the results.}
     } 
    \label{fig:pinn}    
  \end{figure}

 \vspace{-1.0em} 
\subsection{Model Performance with Physics}
This section provides a comparison of the performance of the models with (i.e., PI-GDNN and PI-EVGNN) and without (i.e., GDNN and EVGNN) the use of the physics loss in \eqref{eq:pinnloss}.
Figure \ref{fig:pinn} presents a distribution plot of the Mean Absolute Errors (MAE) for the bus voltage predictions for the four models when applied to the test dataset consisting of N$-1$ test scenarios for all the test networks. For the GD-based models, the MAE values indicate a significantly better performance of the GDNN over the PI-GDNN model with lower average error and a narrower variance. This trend is more evident as the number of buses increases. The average MAE recorded in the case of the 24-bus system is 0.0215 for the GDNN and 0.0276 for the PI-GDNN. A similar trend was observed for the Guided-Dropout NNs, in the 118-bus network, where the average MAE is 0.0342 for the GDNN and 0.3900 for the PI-GDNN.
The reverse is, however, true for the GNN-based models, which indicates a significant improvement in performance as network physics is captured in the model training. In this case, the 24-bus system shows average errors of 0.0489 for the EVGNN and 0.0424 for the PI-EVGNN models, while for the 118-bus network, we obtain MAE of 0.0566 for the EVGNN and 0.0368 for the PI-EVGNN models.

Comparing the best performing framework (i.e., physics-informed versus plain NN) for each model (i.e., Guided-Dropout versus GNN) for the case of the 118-bus network, the GDNN has an average MAE of 0.0342 while the PI-EVGNN has an average of 0.0368. For the 24-bus network, an average error of 0.0215 for the GDNN and 0.0425 for the PI-EVGNN model are obtained. This is a result of the larger variance of the PI-EVGNN models as compared to the GDNN models hence resulting in a slightly better performance of the GDNN as compared to the PI-EVGNN.
In the following, the analysis continues with only the GDNN and PI-EVGNN models as they provided the best performance in each case. 

 \vspace{-0.8em}
\subsection{Voltage and Line Loading Predictions}\label{sec: scatter}

\begin{figure*}[!th]
\centering
\vspace{-0.0cm}
    \subfloat[{N$-1$}\label{fig:v_mae1}]{%
     \scalebox{0.75}{ \includegraphics[width = 0.42\textwidth]{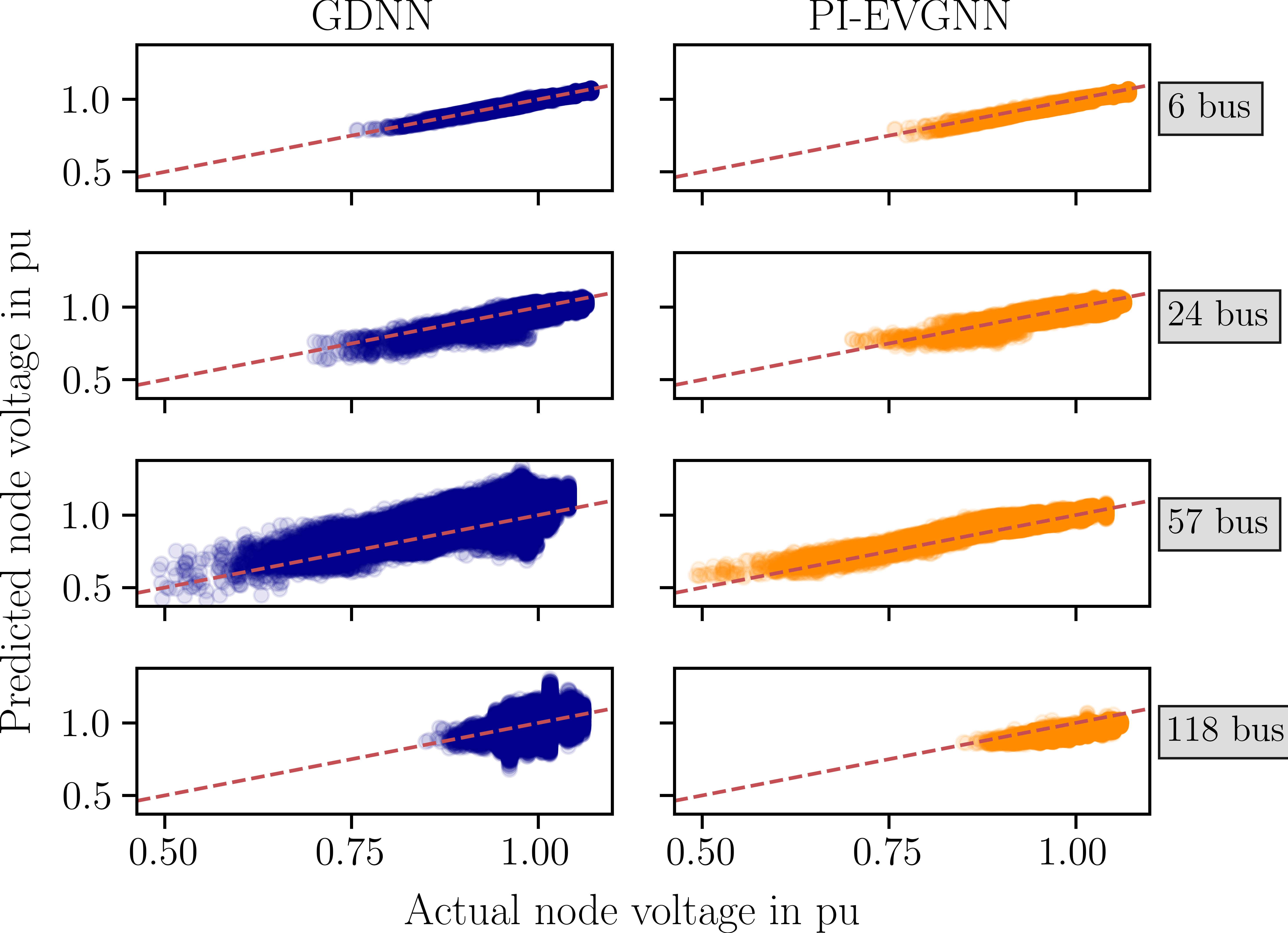}}
    } \hspace{-0.35em} \rulesep
     \subfloat[N$-2$\label{fig:v_mae2}]{%
     \scalebox{0.75}{ \includegraphics[width = 0.42\textwidth]{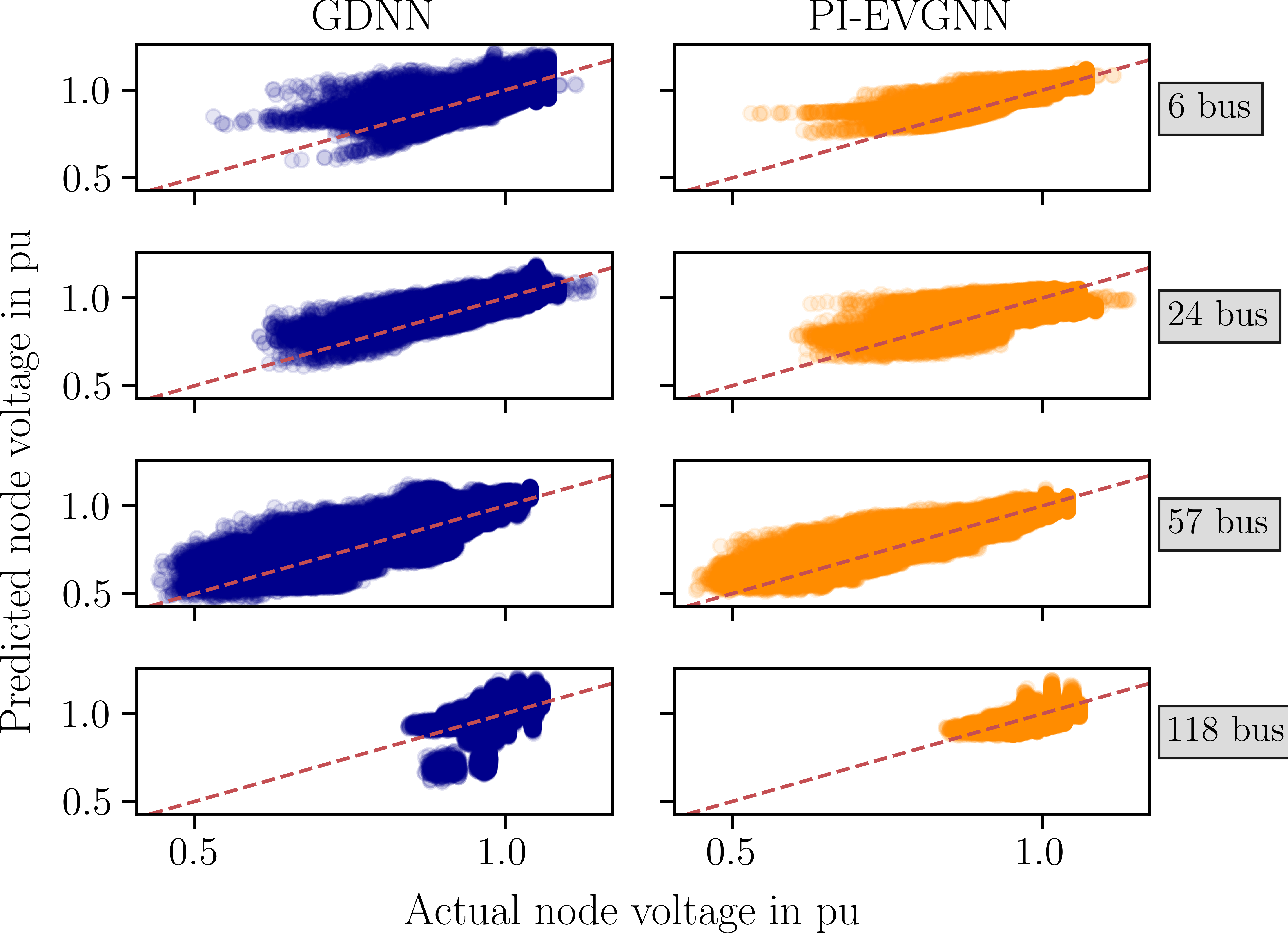}}
    } \hspace{-0.35em} \rulesep
     \subfloat[N$-3$\label{fig:v_mae3}]{%
     \scalebox{0.75}{ \includegraphics[width = 0.42\textwidth]{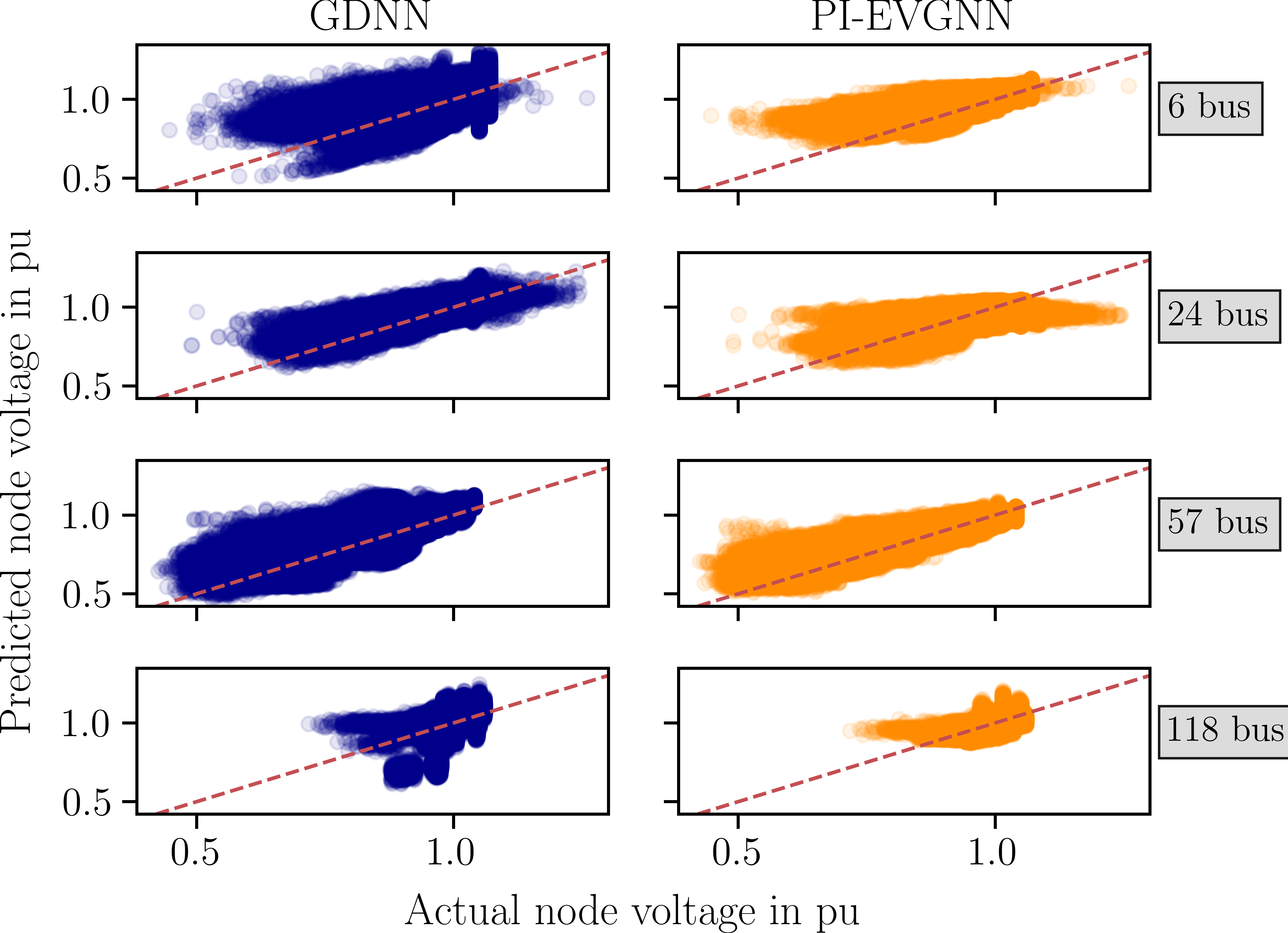}}
    } 
     \caption{ Scatter plots illustrating the relationship between actual and predicted values of voltage magnitude for GDNN and PI-EVGNN models across N-1, N-2, and N-3 contingency scenarios.} \vspace{-0.5em}
    \label{fig:v-mae}
  \end{figure*}

 \begin{figure*}[ht]
\centering
\vspace{-0.0cm}
\hspace{-0.55em}
    \subfloat[{N$-1$}\label{fig:l_mae1}]{%
     \scalebox{0.75}{ \includegraphics[width = 0.42\textwidth]{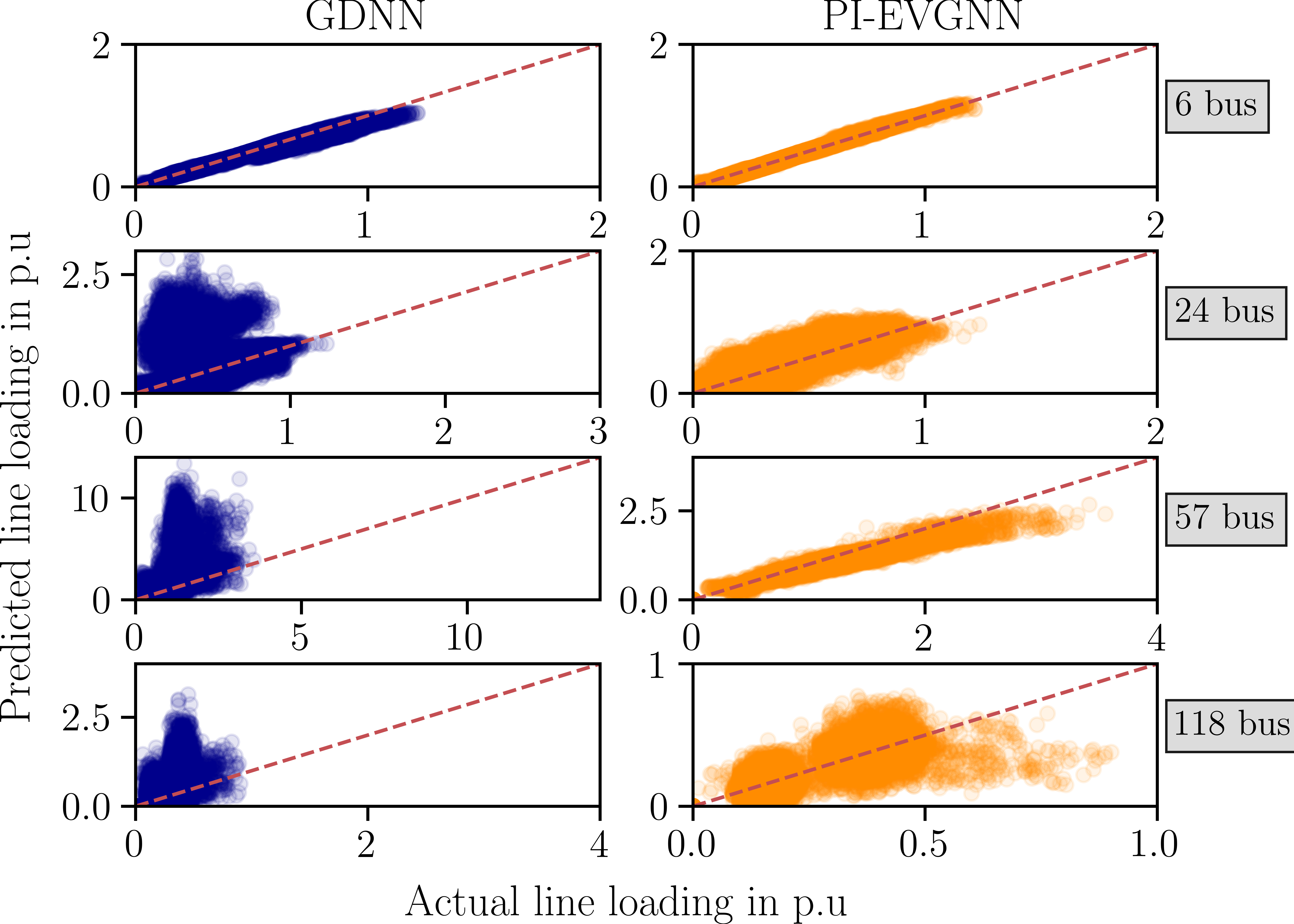}}
    }\hspace{-0.35em} \rulesep
     \subfloat[N$-2$\label{fig:l_mae2}]{%
     \scalebox{0.75}{ \includegraphics[width = 0.42\textwidth]{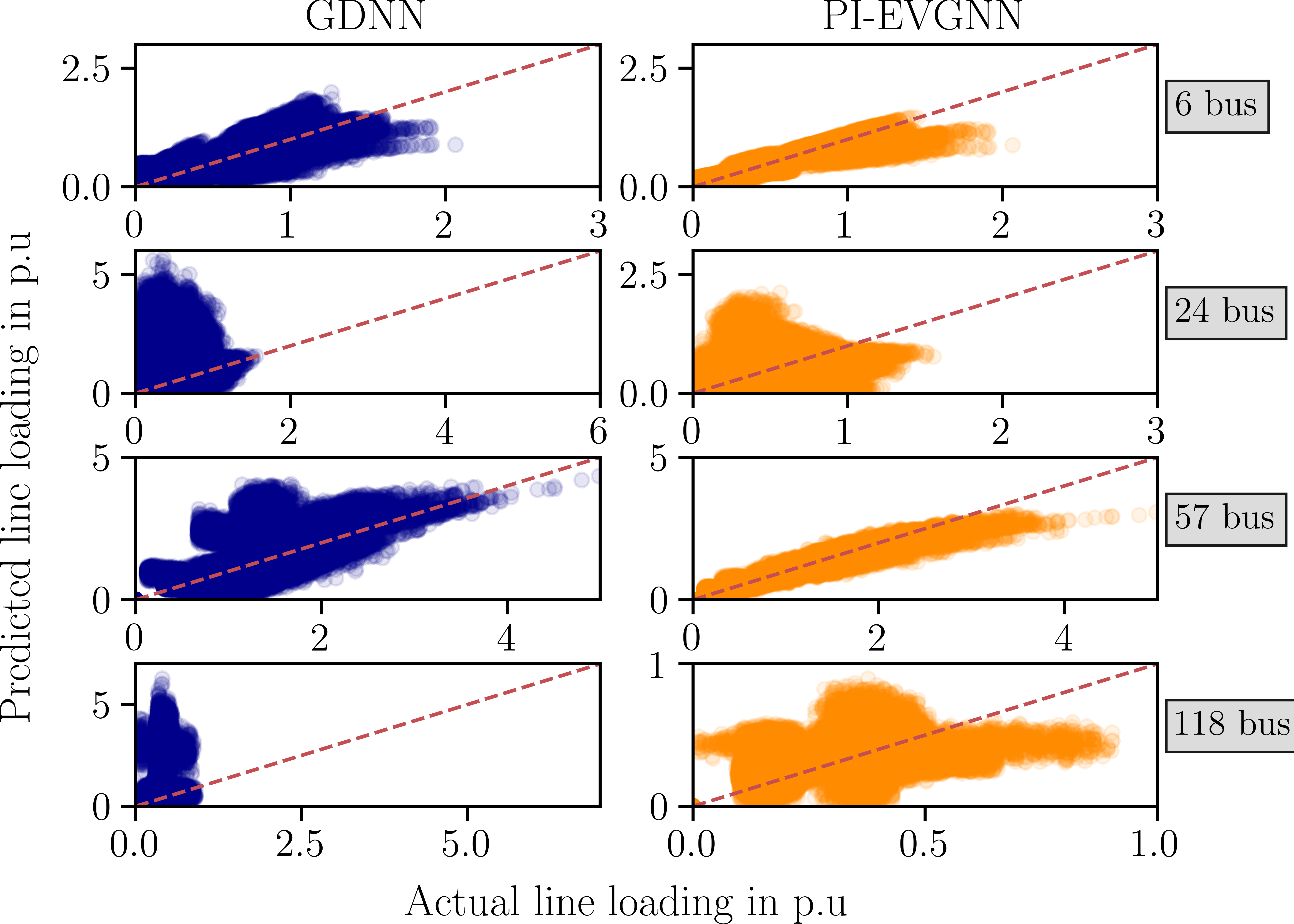}}
    } \hspace{-0.35em}\rulesep
     \subfloat[N$-3$\label{fig:l_mae3}]{%
     \scalebox{0.75}{ \includegraphics[width = 0.42\textwidth]{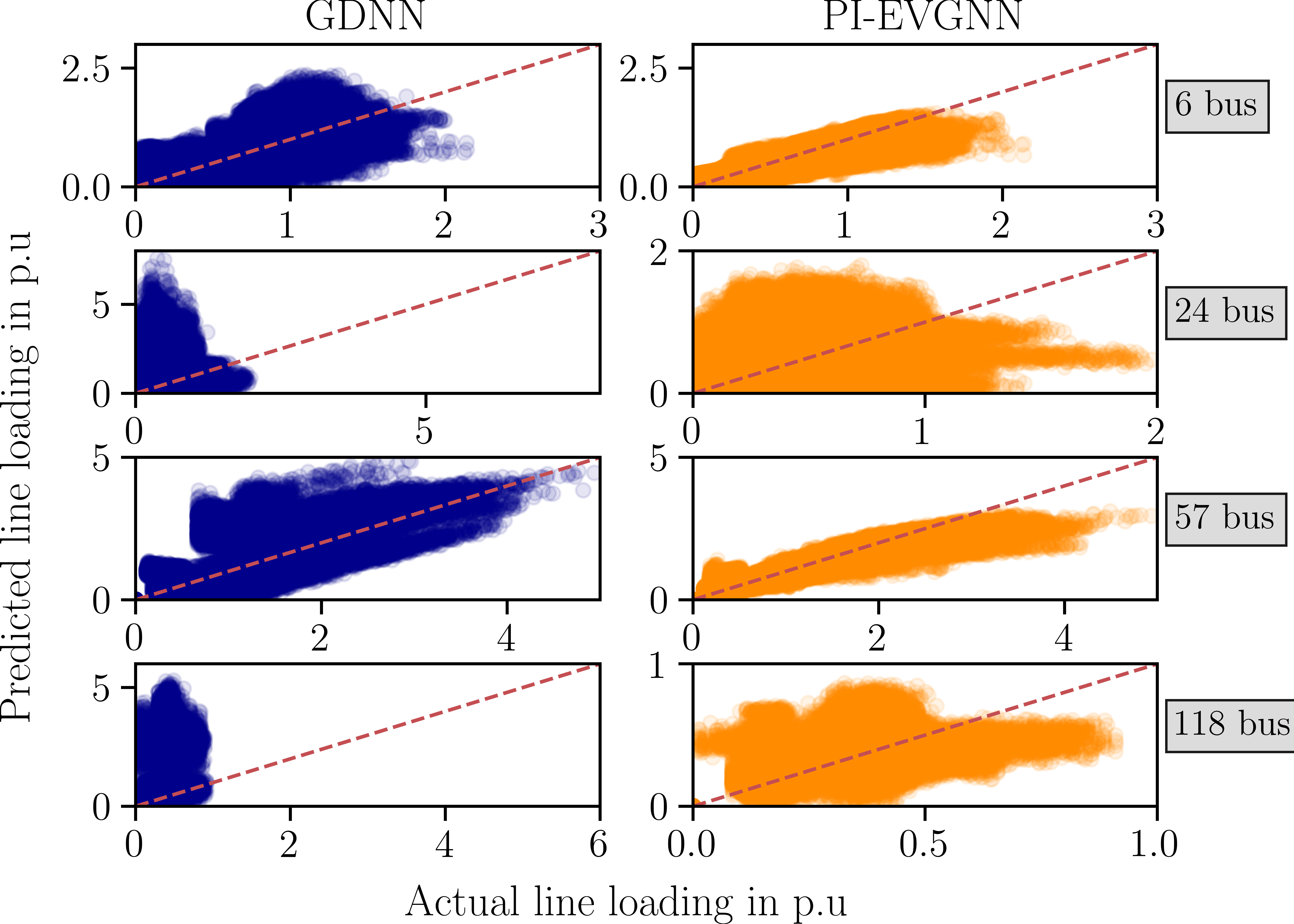}}\hspace{-0.35em}
    } 
     \caption{ Scatter plots illustrating the relationship between actual and predicted values of line loading for GDNN and PI-EVGNN models across N-1, N-2, and N-3 contingency scenarios. {\footnotesize \it Note that different axes have been used for each plot to provide a clearer representation of the results.}}      
    \label{fig:l-mae}
    \vspace{-1.5em}
  \end{figure*}  
  
{\color{black}
We further analyze the prediction performance of the GDNN and PI-EVGNN models by comparing the variance of the predictions from the result of the power flow solver.
Figures \ref{fig:v-mae} and \ref{fig:l-mae} present plots of the predicted against the actual voltage and line power flow for the GDNN and the PI-EVGNN models for N$-1$, N$-2$ and N$-3$ test scenarios. 
The diagonal line indicates equal actual and predicted values. As one can see, there is significant variation in the scatter around the diagonal line between the N$-1$ scenarios and the N$-2$ and N$-3$ scenarios in both  Figs.~\ref{fig:v-mae} and \ref{fig:l-mae}. Points are more concentrated along the diagonal line in the N$-1$ scenarios (i.e. predictions are closer to the actual values) compared to the N$-2$ and N$-3$ cases. This is exactly what we expect, as long as we have trained our models correctly, since the NNs have seen similar N$-1$ topologies in their training, but they have never encountered any N$-2$ or N$-3$ topologies. We remind the reader that the models have on purpose never encountered any N$-2$ or N$-3$ topologies before: our goal with the ML methods we propose is that they learn to provide a good enough estimate of the nodal voltages in any N$-2$ and N$-3$ cases while only learning from the N$-1$ cases, thus avoiding to solve computationally heavy combinatorial problems.

For the voltage predictions in the N$-2$ and N$-3$ cases shown in Figs. \ref{fig:v_mae2} and \ref{fig:v_mae3}, both the GDNN and PI-EVGNN models indicate no significant bias. Overall, predictions are uniformly distributed along the diagonal. 

For the line flow predictions in Fig.~\ref{fig:l-mae}, please notice that the PI-EVGNN has up to 5x smaller axis scales than the GDNN; we acknowledge that this affects perception, but if we maintained the same axis scales, the PI-EVGNN would appear as extremely accurate compared to the GDNN, and the reader would miss the finer details of the PI-EVGNN's performance. 
Inspecting now  Figs. \ref{fig:l_mae2} and \ref{fig:l_mae3}, the GDNN model shows a high upward bias indicating that the line flows are overestimated in many cases. On the other hand, the PI-EVGNN model indicates a downward bias of the points, implying that it underestimates the line flows in some instances, a trend that is very similar for all test networks. Overall, however, the GDNN model shows significantly higher deviations, producing estimates with up to five times higher values than the actual line flows, such as in the case of the N$-3$ test cases for the 24-bus and 118-bus networks. 
In contrast, the worst performance of PI-EVGNN is in the prediction of the active power flow values for the N$-3$ case of the 24-bus network . Overall, however, the PI-EVGNN demonstrates a much better performance in predicting the line flow values compared to GDNN.
}

\vspace{-0.5em}
\subsection{Screening Capability for Critical Voltage and Line Loading }\label{sec: screen}

{\color{black}
Moving from regression tasks, which we discussed in the previous section, in this section we assess the GDNN and PI-EVGNN models in classification tasks. Our goal is to assess their capability to screen critical and non-critical scenarios, i.e. identify undervoltages and line overloading. For the line overloading, besides GDNN and PI-EVGNN, we also use DC Power Flow as a comparison benchmark. DC Power Flow (DCPF) is a common method to approximate power flow results, and screen critical contingencies, substantially speeding up the AC Power Flow solution. Here, the AC Power Flow results are considered as the ground truth: the predictions GDNN, PI-EVGNN, and DCPF are considered correct if they agree with the AC Power Flow results.

We use the recall metric as a means of assessment, as it represents the percentage of correctly identified predictions per class. Considering that our focus is to predict violations, we define as ``positive'' the class of operating points that lead to violations after a contingency (violations are bus over- and undervoltages, and line overloadings), and as ``negative'' the class where there are no violations. As a result, ``true positive'' means that the NN has correctly classified an operating point that leads to a violation (after an N$-1$, N$-2$, or N$-3$ contingency, depending on the test), while ``false negative'' means that the NN has incorrectly classified an operating point as safe, while it truly leads to a violation after a contingency. The sum of ``true positive'' and ``false negative'' collects all operating points and contingencies that truly lead to a violation. In that respect, recall provides insights into the capability of the model to detect critical scenarios.  
}
The recall rate is defined as:
\begin{align}\nonumber
    \text{Recall} = \frac{\text{True Positive}}{\text{True Positive}+\text{False Negative}} \times 100
\end{align}

\begin{figure*}[th!]
\centering
\vspace{-0.0cm}
\hspace{-0.55em}
    \subfloat[{N$-1$}\label{fig:v_rec1}]{%
     \scalebox{0.60}{ \includegraphics[width = 0.42\textwidth]{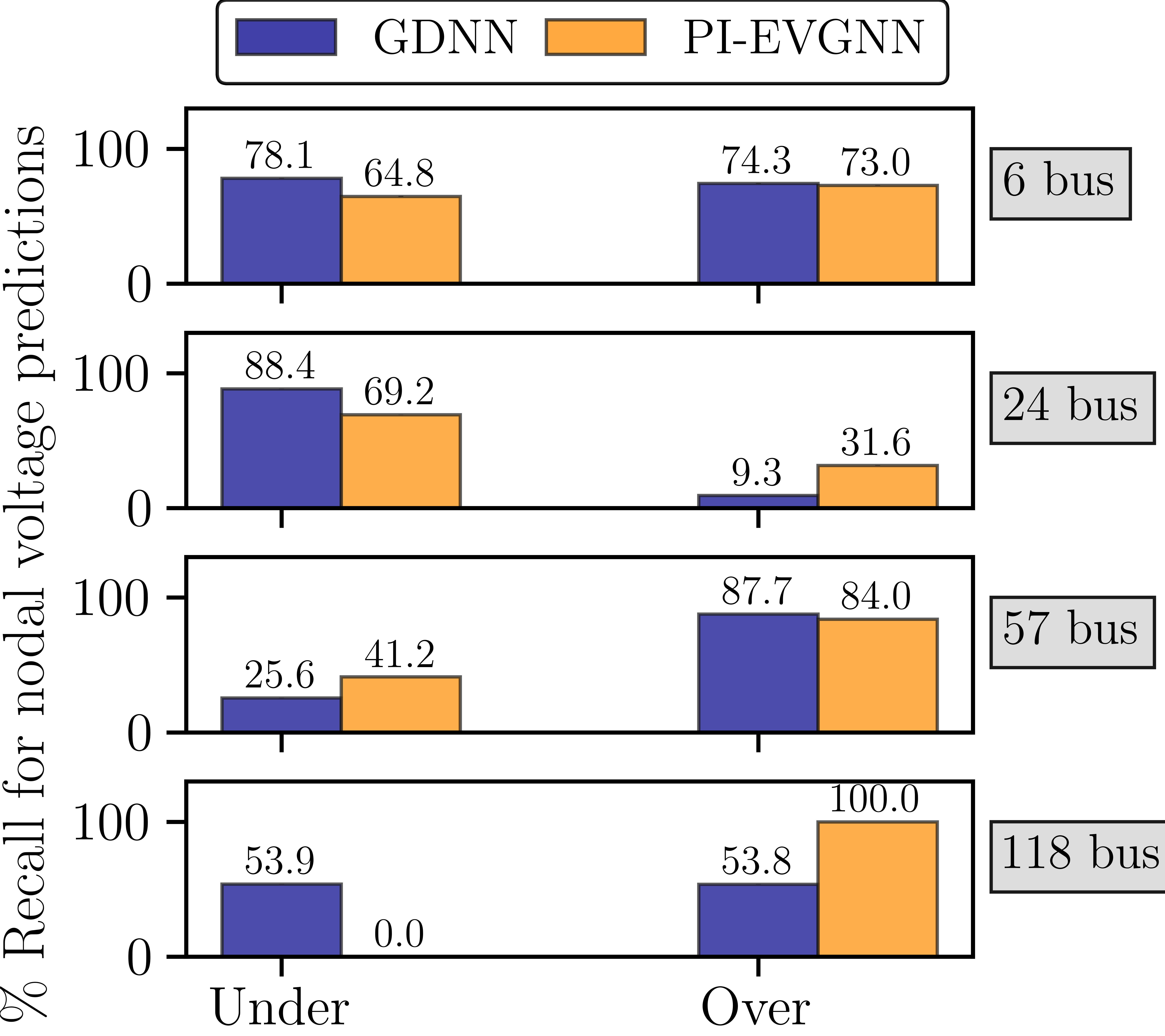}}
    }\hspace{-0.35em} \rulesep
     \subfloat[N$-2$\label{fig:v_rec2}]{%
     \scalebox{0.60}{ \includegraphics[width = 0.42\textwidth]{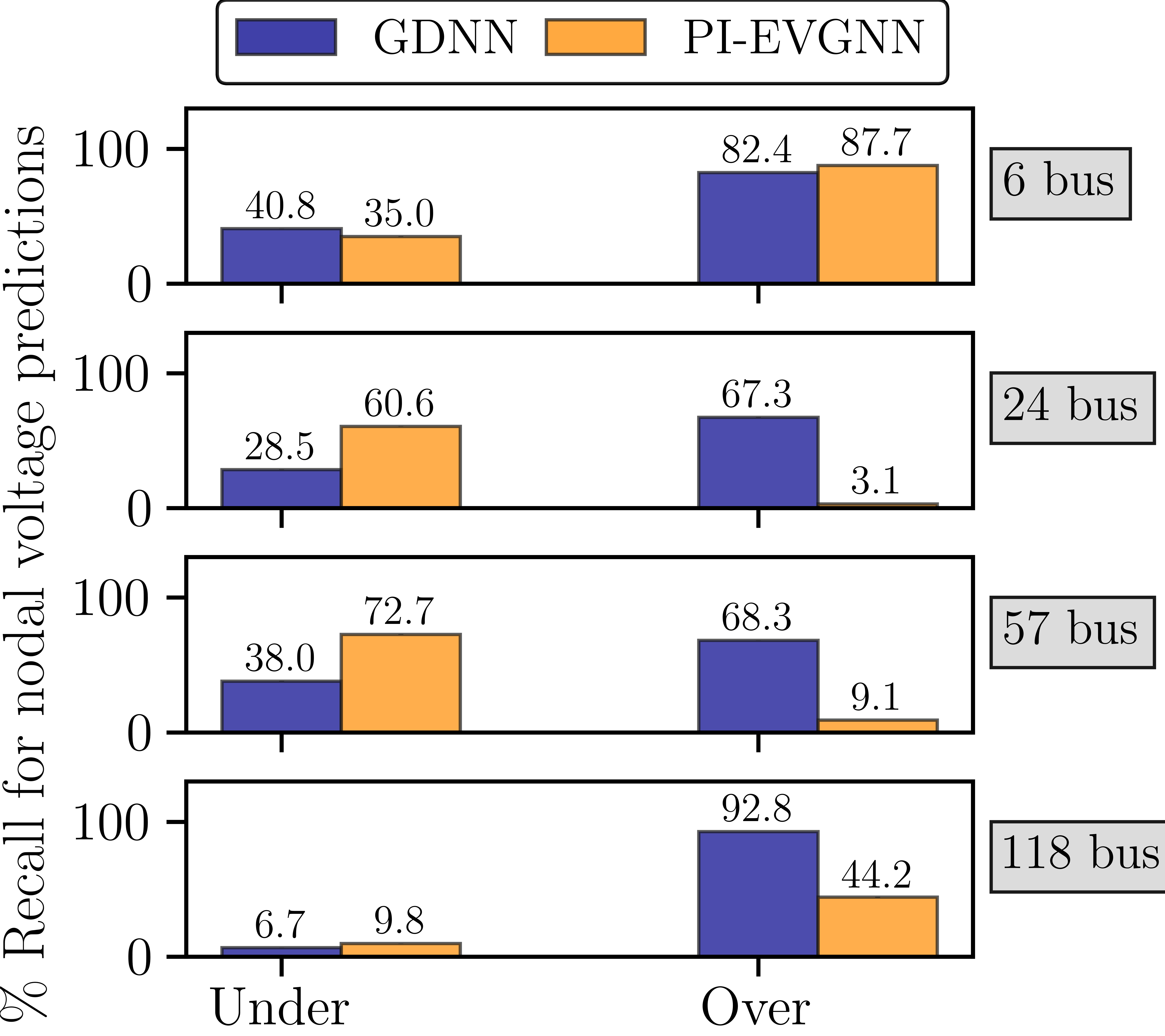}}
    } \hspace{-0.35em}\rulesep
     \subfloat[N$-3$\label{fig:v_rec3}]{%
     \scalebox{0.60}{ \includegraphics[width = 0.42\textwidth]{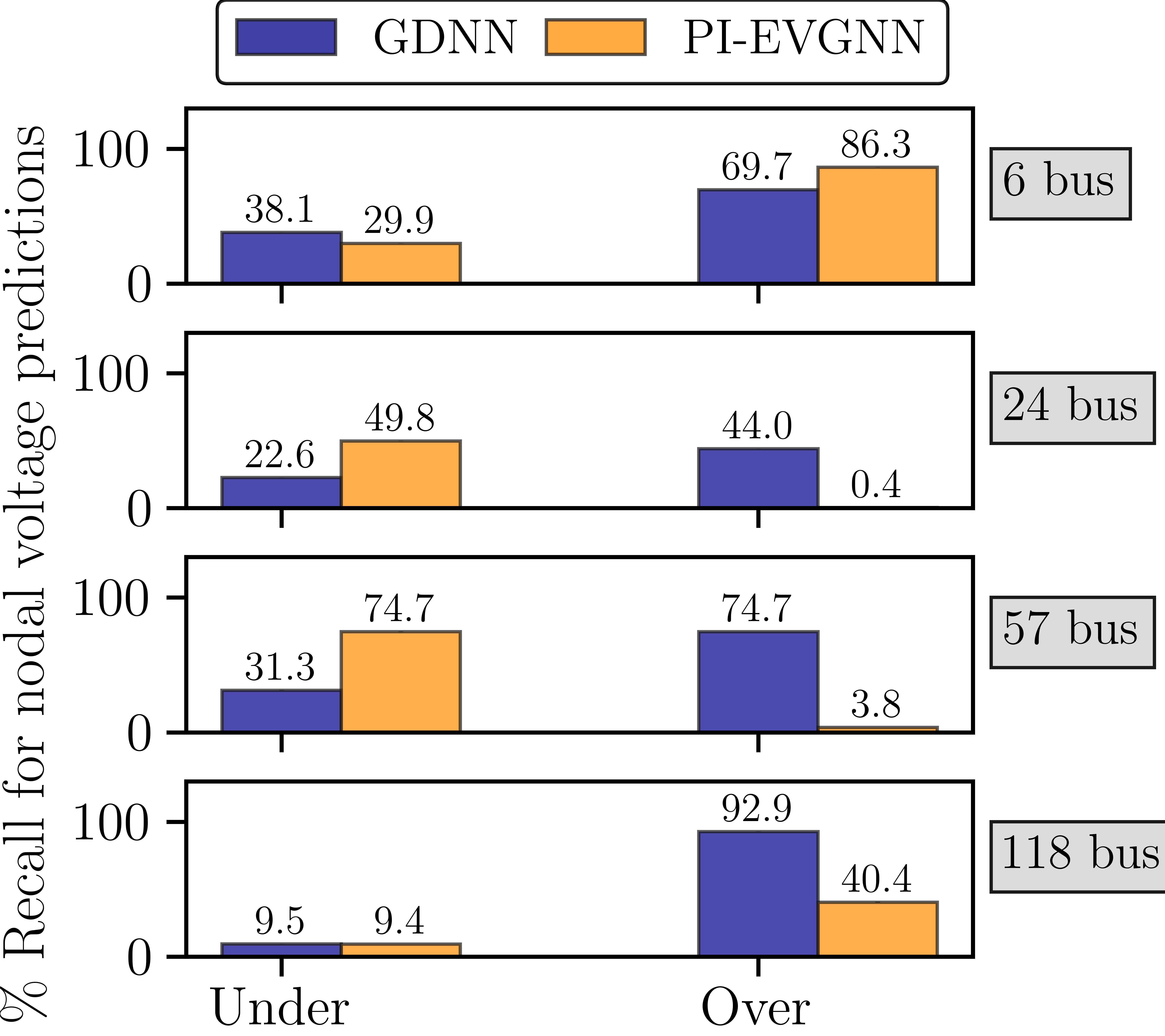}}\hspace{-0.35em}
    } 
     \caption{  Average recall rate for the prediction of nodal under and over voltage by GDNN and PI-EVGNN models across N-1, N-2, and N-3 contingency scenarios. The normal voltage range is set at $0.92 - 1.02$. } \vspace{-0.5em}
    \label{fig:v-pre}
  \end{figure*}

  \begin{figure*}[!ht]
\centering
\vspace{-0.0cm}
\hspace{-0.55em}
    \subfloat[{N$-1$}\label{fig:l_rec1}]{%
     \scalebox{0.65}{ \includegraphics[width = 0.42\textwidth]{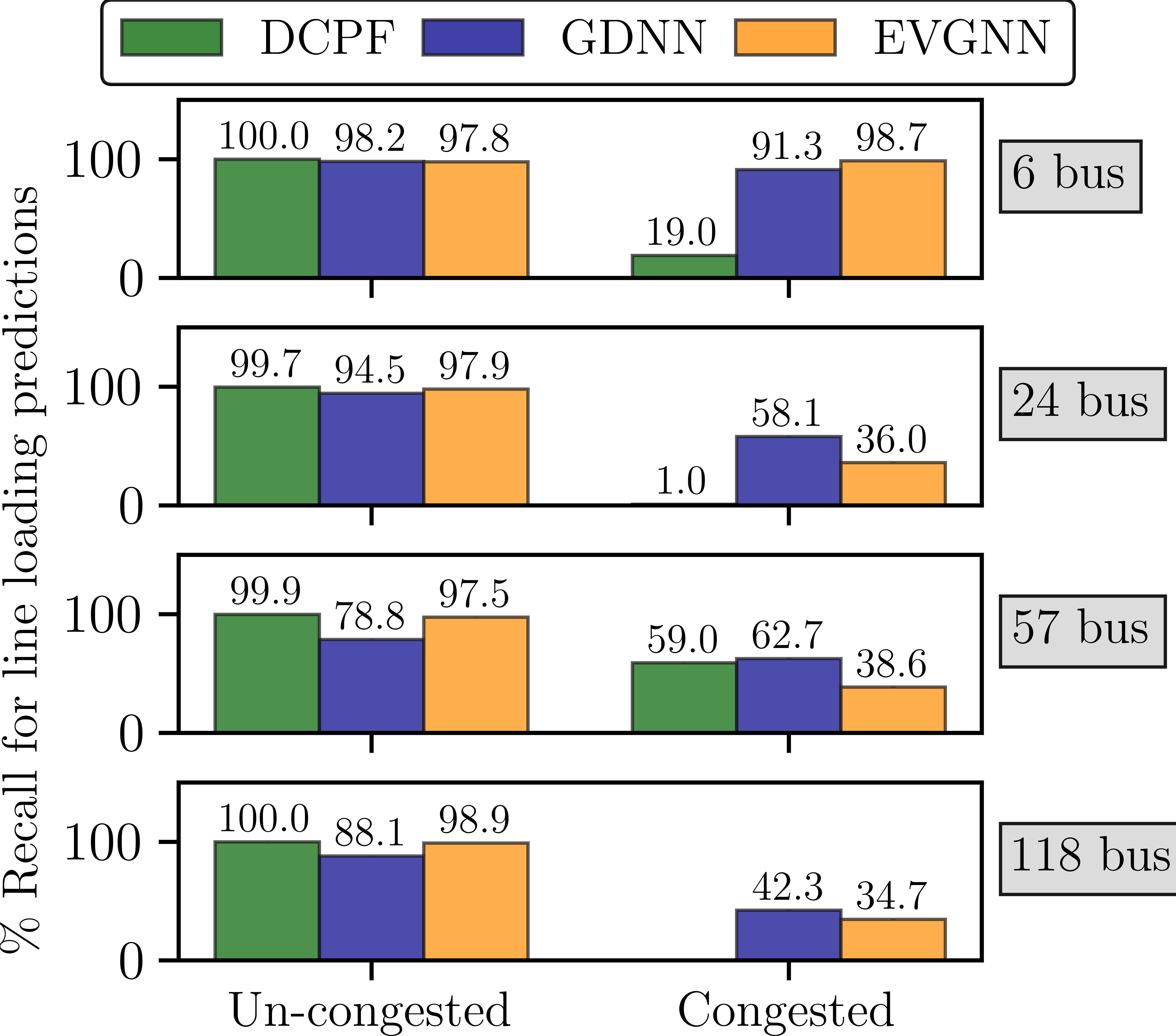}}
    }\hspace{-0.35em} \rulesep
     \subfloat[N$-2$\label{fig:l_rec2}]{%
     \scalebox{0.65}{ \includegraphics[width = 0.42\textwidth]{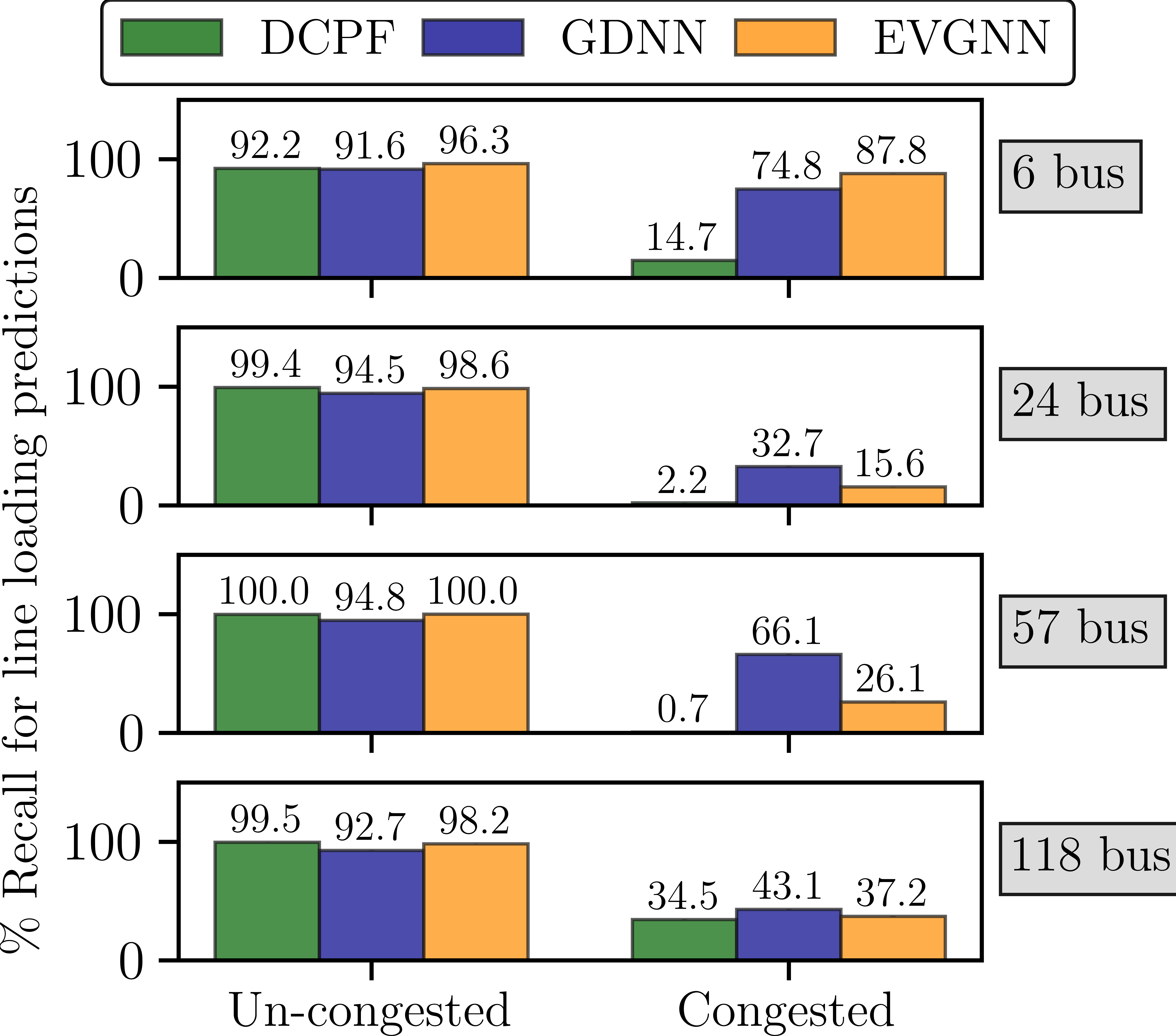}}
    } \hspace{-0.35em}\rulesep
     \subfloat[N$-3$\label{fig:l_rec3}]{%
     \scalebox{0.65}{ \includegraphics[width = 0.42\textwidth]{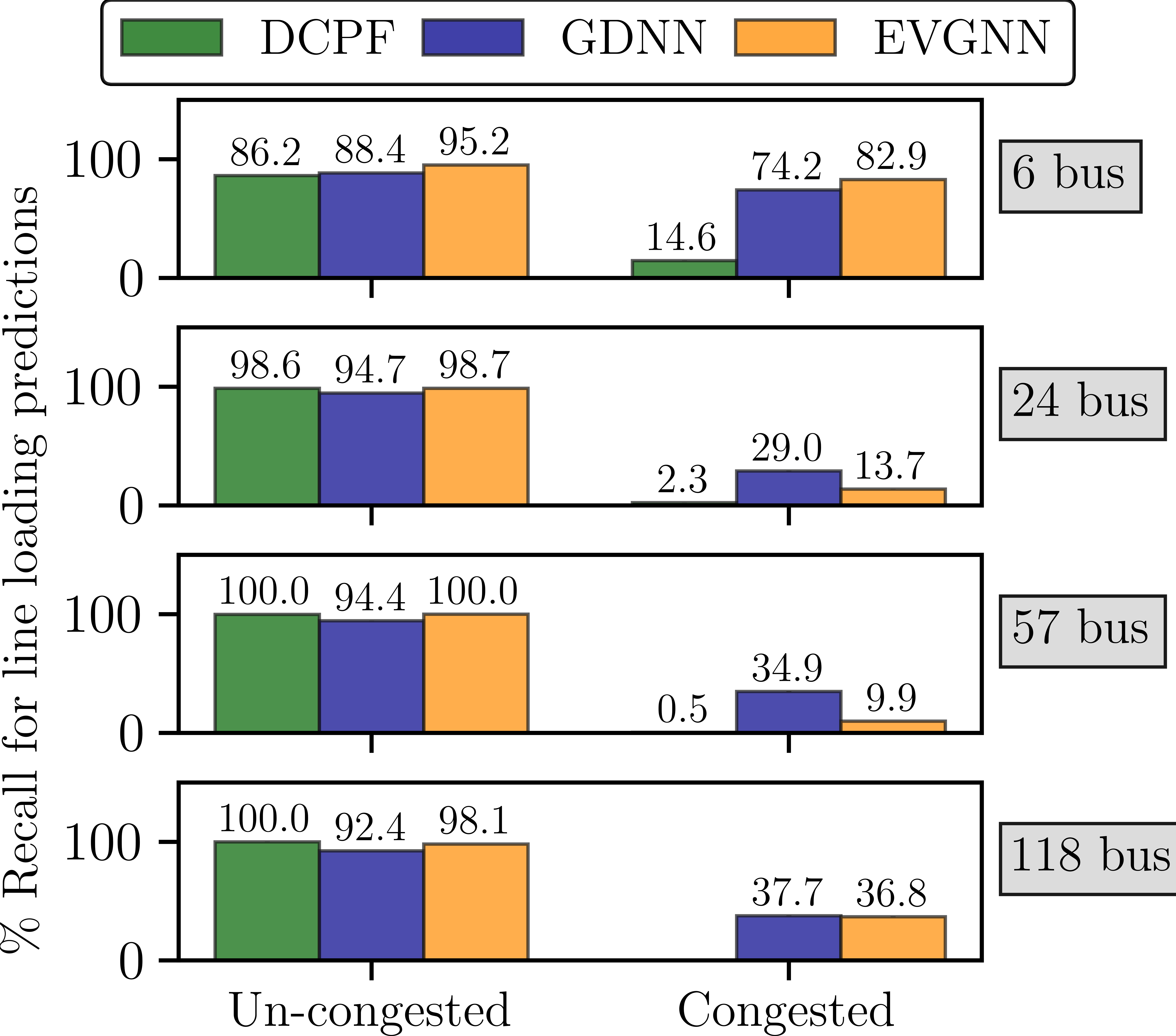}}\hspace{-0.35em}
    } 
     \caption{ Average recall rate for the prediction of line loading by the baseline DCPF, GDNN and PI-EVGNN models across N-1, N-2, and N-3 contingency scenarios.} \vspace{-1.3em}
    \label{fig:l-pre}
  \end{figure*}

A recall rate equal to 100\% denotes that the classifier (here GDNN, PI-EVGNN, or DC Power Flow) has detected all critical scenarios. A recall rate equal to 0\% denotes that the classifier has not detected any of the critical scenarios. In Figs. \ref{fig:v-pre} and \ref{fig:l-pre}, we compare the screening capability for the GDNN, PI-EVGNN, and the DC Power Flow (DCPF) models using the average recall rates. 
Figure \ref{fig:v-pre} presents the recall rates for the nodal voltage predictions for N$-1$, N$-2$ and N$-3$ test data. Here, we compare the GDNN and PI-EVGNN with the ground truth without comparing with the DCPF, as DCPF assumes constant voltages across all buses in its approximation. We use the DCPF as a comparison benchmark in the line overloading predictions in the next figure.

{\color{black} The performance of the NN-based models varies with the size of the test network. For the N$-1$ undervoltage scenarios, GDNN and PI-EVGNN both achieve their best performance on the $6-$bus network; GDNN has its worst performance on the $57-$bus, and PI-EVGNN on the $118-$bus network. For the N$-1$ overvoltage scenarios, on the other hand, both models have a better performance achieving a recall rate above $70\%$ recall rate in the majority of the cases.

For the unseen N$-2$ and N$-3$ scenarios, PI-EVGNN can correctly identify the undervoltage scenarios with a rate above 50\% except for the 118-bus network (see Figs. \ref{fig:v_rec2} and \ref{fig:v_rec3}) in almost all cases. GDNN records averages lower than 40\% in all undervoltage cases.
In contrast, for overvoltages, GDNN has recall rates above $65\%$ in almost all cases, a result synonymous with its overestimation tendency that we described in the previous section. PI-EVGNN, instead, demonstrates a substantial reduction in overvoltage recall rates, with values observed to be below $20\%$ for the $24-$bus and $57-$bus test systems, synonymous with its tendency to underestimate, as we observed in the previous section.

The screening capability for the line congestion is shown in Fig. \ref{fig:l-pre}. Here, we observe that the DCPF, which is a common approximation to speedup these type of computations, has a significantly lower performance compared to the NN-based models, making it often unsuitable to approximate critical scenarios for line overloading. The GDNN generally outperforms the PI-EVGNN in detecting line congestion, however, its average recall rate falls short of 50\% in all but the 6-bus network for most N$-1$, N$-2$ and N$-3$ test scenarios. This trend may not necessarily be attributed only to the prediction performance of the models but to the nature of the training data, which contained unbalanced scenarios for the different classes.} 

{
The results both in Figs.~\ref{fig:v-pre}~and ~\ref{fig:l-pre} speak about the need for highly representative data both during training and testing. Here, we have used up to 450'000 test datapoints for each test system (converged power flows for 1'000 operating points $\times$ $>$1'500 sample topologies (N$-1$, N$-2$, N$-3$)). 
Still, this number may not have vastly represented the potential operating regions and could result in unbalanced classes. 
Generating data that result in balanced datasets across classes and capture the true distribution of the underlying phenomena is an art that requires its own dedicated scientific work. In \cite{Thams_database, VENZKE2021106614}, we discussed these issues and ways to create good-quality datasets for power system security assessment, but further work is necessary. This is an object of our future work.
}


\vspace{-1.0em}
\subsection{Computational Efficiency}\label{sec: times}
 As discussed in the introduction, the potential of NN models to result in substantially reduced computation times is one of the main motivations for their application. In this section, we compare the computation time required to get a prediction from the DC Power Flow (DCPF), AC Power Flow (ACPF; employing the traditional Newton-Raphson method), the GDNN and the PI-EVGNN models. The ACPF is considered the current benchmark in terms of accuracy for assessing critical scenarios. Due to its heavier computational needs, for several decades the DCPF has been used as a surrogate to predict the critical contingencies. And, indeed, considering the limited computational resources of the past, DCPF could offer a non-negligible speedup compared to ACPF. In the previous section, we found that GDNN and PI-EVGNN are clearly more accurate in their critical contingencies prediction compared with DCPF. Our goal here is to assess how GDNN and PI-EVGNN compare with ACPF and DCPF in terms of computing time.  

First, we consider execution time only. Afterward, we consider the total computation time, which includes the dataset generation and the training time for the GDNN and PI-EVGNN models. 

\begin{figure}[!tb]
    \centering
    \includegraphics[width=.8\linewidth]{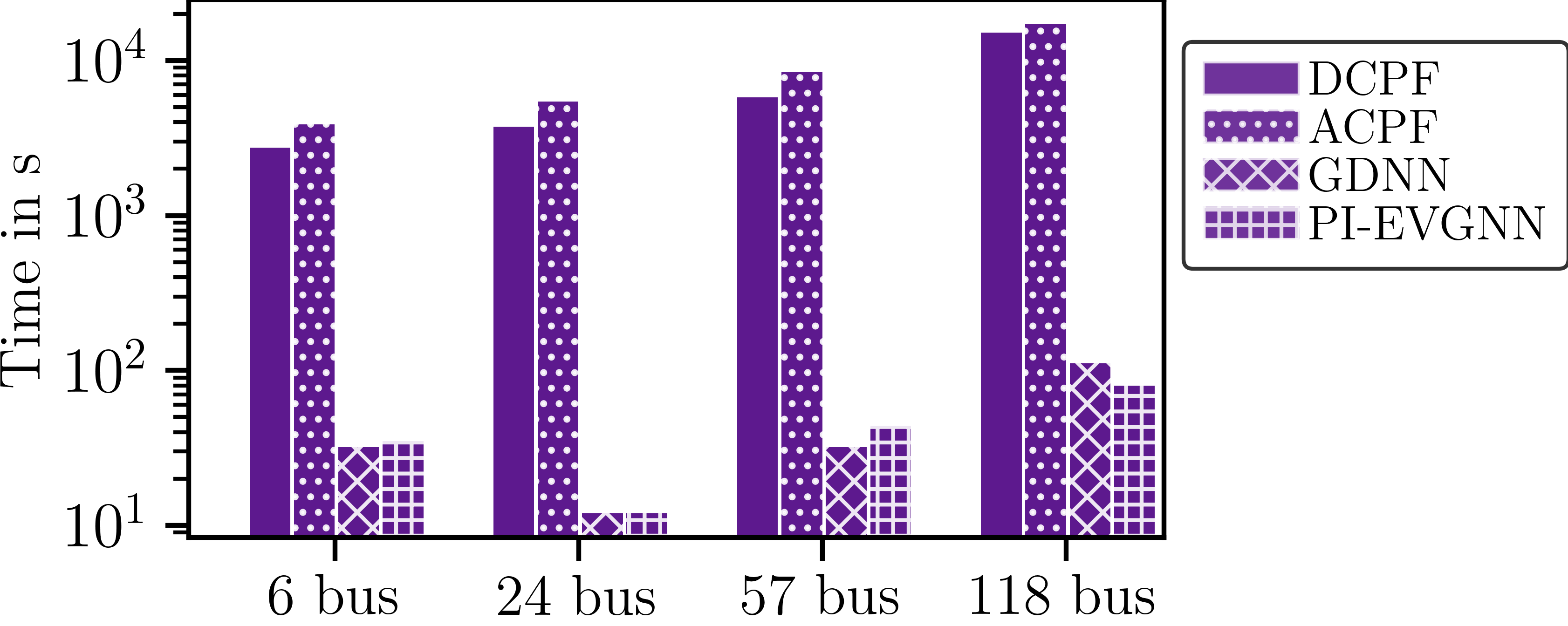}
    \vspace{-.5em}
    \caption{Average time taken to test 100,000 N$-3$ test scenarios by ACPF, DCPF, pre-trained GDNN, and pre-trained PI-EVGNN models across the test networks. Please notice the logarithmic scale. }
    \label{fig:test_time}
					
\vspace{-1.5em}
\end{figure}

Fig.~\ref{fig:test_time} presents the average time taken to assess 100,000 N$-3$ scenarios with DCPF, ACPF, GDNN, and PI-EVGNN for the 4 different test systems: the 6-bus, 24-bus, 57-bus, and 118-bus. Please notice the logarithmic scale on the y-axis, which makes the bar length disproportionate. From Fig.~\ref{fig:test_time}, it becomes clear that GDNN and PI-EVGNN are 100-400 times faster than their DC and AC Power Flow counterparts. In other words, while GDNN and PI-EVGNN need approximately 1.5 minutes to assess 100,000 scenarios in the 118-bus system, DCPF and ACPF need almost 5 hours for the same task. This can indeed be a \textcolor{black}{game-changer} for real-time security assessment, and it shows that a good implementation of NN algorithms can allow for a much more comprehensive real-time security assessment compared to the current state of the art. On top of that, once trained, GDNN and PI-EVGNN can run on a simple laptop, not requiring any advanced or expensive computing resources. 

\begin{figure}[!tb]
    \centering
    \includegraphics[width=.9\linewidth]{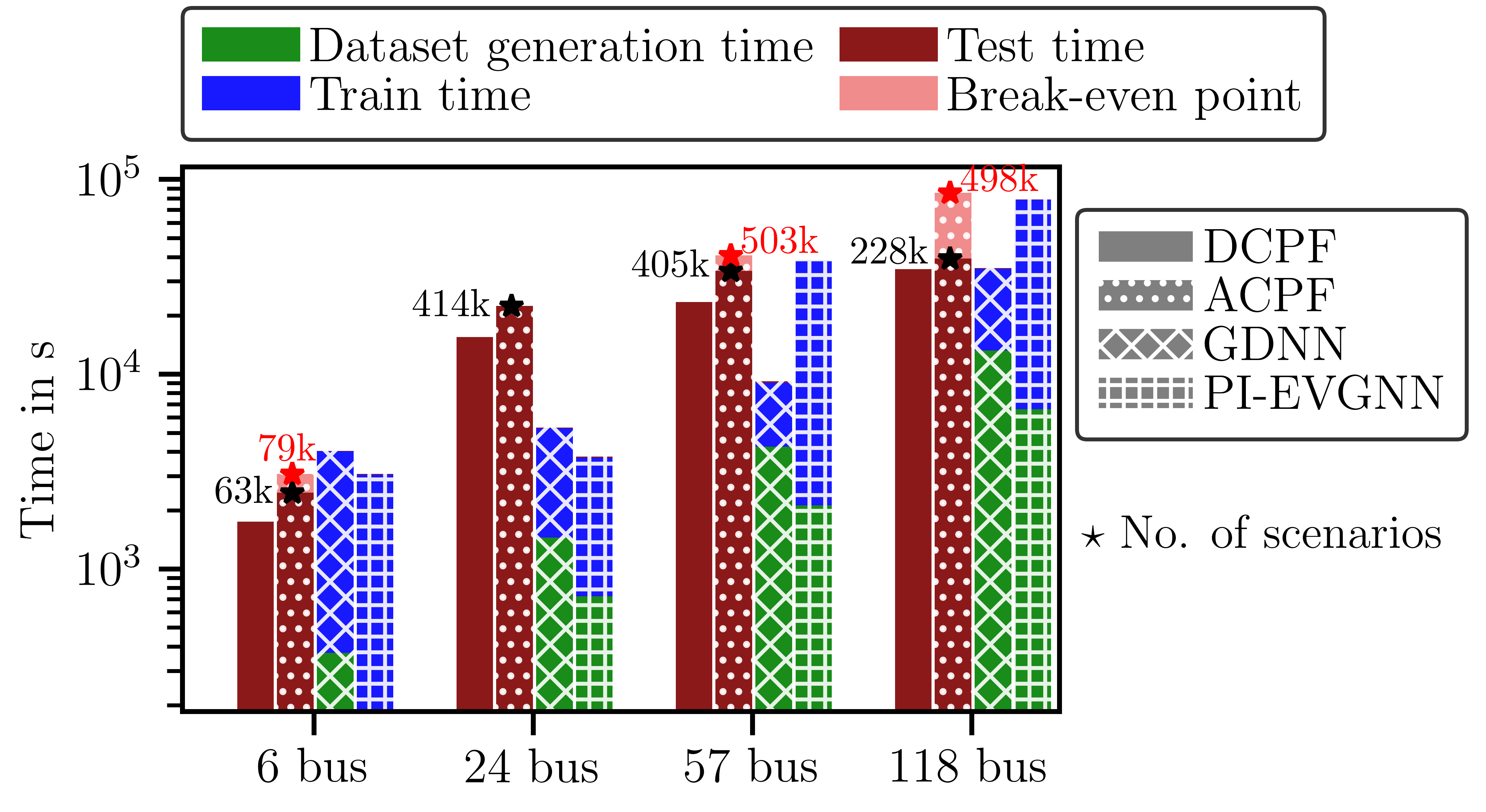}
    \vspace{-1em}
    \caption{Total computation times for the ACPF, DCPF, GDNN and PI-EVGNN models to assess N$-3$ scenarios across the test networks. The black $\ast$ indicates the number of scenarios considered during testing. The red {\color{red}$\ast$} indicates the break-even point in total speed between ACPF and PI-EVGNN. Please notice the logarithmic scale.}
    \label{fig:comp_time}
\vspace{-1.0em}
\end{figure}

As it can be deduced, the computational advantage of GDNN and PI-EVGNN in real-time derives from shifting a substantial part of the computation time offline; this is the time required for generating the training dataset and training the NN models. In Fig. \ref{fig:comp_time}, we present the total computation time, again for the DCPF, ACPF, GDNN, and PI-EVGNN, for all 4 test systems. This total time is composed of (i) the dataset generation time denoting the time used to generate the N$-0$ and N$-1$ training scenarios, (ii) the model training time, and (iii) the model test time (this is the execution time we discussed in Fig.~\ref{fig:test_time}). Please notice again the logarithmic scale, as it makes the bar length not proportional. Since DCPF and ACPF are purely analytical models, they have zero data generation and training time. In contrast, the test time for the GDNN and PI-EVGNN models is negligible (almost not visible in Fig. \ref{fig:comp_time}) compared to DCPF and ACPF models. The annotations show the number of N$-3$ scenarios tested for each test system. Although we have originally aimed for 1'500'000 test scenarios for each system (1,000 generation and demand scenarios $\times$ (500 N$-2$ + 1'000 N$-3$) contingencies), the size and topology of each system led to a number of them being discarded either because the contingencies led to islanded systems or because the AC power flow did not converge. This resulted in having 63'000 test scenarios for the 6-bus system, 414'000 test scenarios for the 24-bus system, 405'000 test scenarios for the 57-bus system, and 228'000 test scenarios for the 118-bus system. We can see that despite having almost half of the scenarios in the 118-bus system compared to the 57-bus system, the ACPF computation times are almost equal. This is because it is more computationally intensive to compute the ACPF for a larger system.

In terms of total computation speed, we see that GDNN is already faster than DCPF and ACPF in all cases except for the 6-bus case, which has too few scenarios. This means that it is faster to generate the N$-0$ and N$-1$ data, train the GDNN, and assess the 220'000-500'0000 scenarios than to execute the 220k-500k DCPF/ACPF themselves.

About PI-EVGNN, since it includes the underlying physical equations in the training, we see that it requires less training data than the GDNN (green bars are shorter) but it takes a longer training time, despite having 40\% fewer parameters as shown in Table~\ref{tab:trainparameters}. Physics-Informed NN training procedures are not as mature as the standard NN training yet. With the rapidly increasing focus on Physics-Informed Machine Learning across communities, we expect substantial improvements in their training will ensue, similar to the improvements general purpose optimization solvers have experienced in the past.

In terms of the total computation speed, we see that for the 24-bus system PI-EVGNN is already about 10 times faster than DCPF and ACPF. For the larger systems, we see that the break-even point is at approximately 500k scenarios (plotted on a lighter-colored bar above ACPF, and annotated in red: 503k for the 57-bus and 498k for the 118-bus). This translates to approximately 1'000 operating points and 500 contingencies. Considering that the 118-bus has over 800'000 N$-2$ and N$-3$ contingencies (and the 57-bus over 39'000), one can easily deduce that large systems require the testing of hundreds of millions of scenarios, where the solution of GDNN and PI-EVGNN will provide an orders of magnitude faster alternative.

\vspace{-0.5em}
\section{Conclusion} \label{sec:concl}
The capability of Graph Neural Networks to analyze systems with changing topologies is key for problems such as the N-\textit{k} security analysis in power systems.  Our goal in this paper is to create tools that are drastically faster than existing solutions to screen millions of N-\textit{k} scenarios. As soon as they have identified a set of critical scenarios, the AC Power Flow can then be for more detailed analysis.

This paper presents two promising graph-aware and physics-informed Neural Network (NN) models: the Guided-Dropout Neural Network (GDNN), and the Physics-Informed Edge-Varying Graph Neural Network (PI-EVGNN). To the best of our knowledge, this is the first paper that introduces a Physics-Informed Architecture of the GDNN model. Our objective is to avoid the combinatorial explosion of assessing N$-k$ topologies; for example, the 118-bus system has 166 N$-1$ topologies, but it has almost 800'000 N$-3$ topologies. Therefore, we train GDNNs and PI-EVGNNs \emph{only} on the N$-0$ and N$-1$ contingencies and assess their performance for N$-2$ and N$-3$ contingencies, comparing them against the DC and AC Power Flow (DCPF, ACPF).

{\color{black}
Overall, GDNN and PI-EVGNN provided a clearly better performance as compared to the widely adopted DCPF for the line flow violations and showcased high precision and contingency screening capabilities. 
The results demonstrate that the NN models can perform accurate critical contingency screening for N$-2$ and N$-3$, without the necessity of exhaustive re-training across the combinatorial space of N$-k$ configurations and their associated scenarios. 
Overall, the Physics-Informed EVGNN model delivered higher precision accuracy than the GDNN model. While the GDNN presented a tendency to overestimate predictions in many cases, lending it beneficial in some operational scenarios where robustness is required, the PI-EVGNN model shows a narrower distribution of the predictions, providing a closer relation to reality.

In terms of computation speed, the NNs were 100-400 times faster than the AC Power Flow Solution. In other words, both GDNN and PI-EVGNN can assess 100'000 in only 1.5 minutes, while the AC Power Flow required 5 hours. When it comes to the total computation, where the time to generate the training data and the NN training is accounted for, then GDNN is still faster than AC Power Flow for the number of scenarios considered. For the PI-EVGNN, it appears that the break-even point for larger systems is at approximately 500'000 scenarios, a number which is probably small if we consider that for 1'000 potential operating points, we need to assess up to 800 million N$-3$ scenarios in the 118-bus case. Considering the findings in this paper, we think that both GDNN and PI-EVGNN can become excellent candidates for rapid N$-k$ contingency screening in power systems. Our results show that GDNN requires less training time, but PI-EVGNN appears to be more accurate.

Before concluding this paper, we wish to stress that to ensure robust GNN performance, we need to carefully consider the selection of data samples and collocation points used to train the NN models, with a specific focus on achieving training datasets with representative coverage across all operating regions, encompassing both critical and non-critical scenarios. Our future enhancement of the models will ensure that security boundaries separating operation regions are distinctly modeled to ensure balanced datasets are employed during model training. Furthermore, we will focus on enhancing the models' ability to handle worst-case scenarios and provide stronger guarantees for their adoption. Additionally, we will investigate interpretability mechanisms to gain a deeper understanding of how graph-aware neural networks operate for power systems, given their intrinsic nature.
}
\vspace{-0.5em}

\bibliographystyle{IEEEtran}
\bibliography{references}
\end{document}